\documentclass[twocolumn,a4paper,11pt]{article}
\usepackage{amsmath,amssymb,amsfonts,amsthm,makeidx,graphicx}
\usepackage{txfonts}
\usepackage{helvet}
\usepackage{hyperref}
\usepackage{graphics}
\usepackage{rotating} 
\usepackage{graphicx}
\usepackage{amssymb}
\usepackage{amsmath}
\usepackage{xcolor}
\usepackage{mathtools,slashed}
\usepackage{epstopdf}
\usepackage{xspace}  
\usepackage{xspace}

\begin{document}

\newcommand{\ptt}{\ensuremath{p_{\rm{T}}}\xspace}
\newcommand{\de}{\ensuremath{\rm{{\delta \eta}}}\xspace}
\newcommand{\pgev}{\ensuremath{{\rm GeV}/c}\xspace}
\newcommand{\ptev}{\ensuremath{{\rm TeV}/c}\xspace}
\newcommand{\gev}{\ensuremath{{\rm GeV}}\xspace}
\newcommand{\tev}{\ensuremath{{\rm TeV}}\xspace}
\newcommand{\pmev}{\ensuremath{{\rm MeV}}\xspace}
\newcommand{\mev}{\ensuremath{{\rm MeV}/c}\xspace}
\newcommand{\nch}{\ensuremath{\rm{N_{ch}}}\xspace}
\newcommand{\anch}{\ensuremath{\rm{ < N_{ch}} >}\xspace}
\newcommand{\pp}{\ensuremath{\rm p\!+\!p}\xspace}
\newcommand{\ppb}{\ensuremath{\rm p\!+\!Pb}\xspace}
\newcommand{\pbpb}{\ensuremath{\rm Pb\!+\!Pb}\xspace}
\newcommand{\auau}{\ensuremath{\rm Au\!+\!Au}\xspace}
\newcommand{\ncr}{\ensuremath{\rm{CR}}\xspace}
\newcommand{\nica}{\ensuremath{\rm{NICA}}\xspace}
\newcommand{\rhic}{\ensuremath{\rm{RHIC}}\xspace}
\newcommand{\sps}{\ensuremath{\rm{SPS}}\xspace}
\newcommand{\lhc}{\ensuremath{\rm{LHC}}\xspace}
\newcommand{\fair}{\ensuremath{\rm{FAIR}}\xspace}
\newcommand{\alice}{\ensuremath{\rm{ALICE}}\xspace}
\newcommand{\mpdnica}{\ensuremath{\rm{MPD-NICA}}\xspace}
\newcommand{\naexp}{\ensuremath{\rm{NA49}}\xspace} 
\newcommand{\start}{\ensuremath{\rm{STAR}}\xspace}
\newcommand{\ags}{\ensuremath{\rm{AGS}}\xspace}
\newcommand{\qgp}{\ensuremath{\rm{QGP}}\xspace}
\newcommand{\s}{\ensuremath{\rm{S}}\xspace}
\newcommand{\q}{\ensuremath{\rm{Q}}\xspace}
\newcommand{\ba}{\ensuremath{\rm{B}}\xspace}
\newcommand{\KaonToPionPlus}{\ensuremath{\rm{K^+/\pi^+}}\xspace} 
\newcommand{\KaonToPionMinos}{\ensuremath{\rm{K^-/\pi^+}}\xspace} 
\newcommand{\LambdaToPi}{\ensuremath{\rm{\Lambda^0/\pi^{\pm}}}\xspace} 
\newcommand{\ALambdaToPi}{\ensuremath{\rm{\overline{\Lambda^0}/\pi^{\pm}}}\xspace} 
\newcommand{\KaonToKaonPM}{\ensuremath{\rm{K^-/K^+}}\xspace} 
\newcommand{\OmegaToOmegaPM}{\ensuremath{\rm{\overline{\Omega}/\Omega}}\xspace} 
\newcommand{\ALambdaLambda}{\ensuremath{\rm{{\overline{\Lambda^0}}/\Lambda^0}} \xspace}  
\newcommand{\qcd}{\ensuremath{\rm{QCD}}\xspace}
\newcommand{\ce}{\ensuremath{\rm{CE}}\xspace}
\newcommand{\gce}{\ensuremath{\rm{GCE}}\xspace}
\newcommand{\mce}{\ensuremath{\rm{MCE}}\xspace}
\newcommand{\hrg}{\ensuremath{\rm{HRG}}\xspace}
\newcommand{\qvdw}{\ensuremath{\rm{QVdW}}\xspace}
\newcommand{\thermal}{\ensuremath{\rm{THERMAL-FIST}}\xspace}

\title{Energy and centrality dependence of thermodynamical observables from multiplicity in \pbpb and \auau collisions}
\author{Francisco Reyes Rodr\'iguez  and Eleazar Cuautle\thanks{ecuautle@nucleares.unam.mx}\\
  {\small  Instituto de Ciencias  Nucleares, Universidad Nacional Aut\'onoma de M\'exico,}\\
  {\small Apartado Postal 70-543, Ciudad de M\'exico 04510,  M\'exico.}
}



\maketitle 
\abstract{
  Using a statistical model, we analyze published multiplicity distributions for identified hadrons produced in heavy ion collisions in the energy range from 2.7-200 \gev. Our analysis enables the prediction of the multiplicity distributions for multi-strangeness hadrons that have not yet been measured in lower-energy experiments. Furthermore, we obtain freeze-out parameters, including temperature, baryon chemical and strangeness potentials, the strangeness suppression factor, and the system radius, as functions of centrality and collision energy.
 Additionally, we computed and discussed the Skewness and its behavior at lower collision energies, highlighting the importance of the freeze-out parameters in determining the liquid-gas phase transition in nuclear matter.
}

\section{Introduction}
\label{intro}

Heavy-ion collisions can produce a strongly interacting medium, which has been studied using Lattice Quantum Chromodynamics (\qcd). This theory predicts the transition between hadronic matter and quark-gluon plasma (\qgp) with a cross-over~\cite{Aoki:2006we,Cheng:2007jq}. We can find a recent theoretical review of the \qcd phase diagram and the connection with experimental observables in Ref.~\cite{Akiba:2015jwa,Du:2024wjm}.
Experimentally, there are efforts to understand the \qgp in \rhic~\cite{BRAHMS:2004adc,PHOBOS:2004zne,STAR:2005gfr,PHENIX:2004vcz}, as well as to characterize it by freeze-out parameters at \lhc~\cite{Floris:2014pta} accelerators. 
One of the most important research areas is looking at the critical endpoint, which can be studied with  event-by-event fluctuation of conserved quantum numbers with real data, 
allowing various analyses, with particular emphasis on those related to strangeness production, as one of the variables proposed to investigate \qgp~\cite{Rafelski:1982pu}. In this direction, the \LambdaToPi and \KaonToPionPlus ratios~\cite{NA49:2004iqm} are examples of analyses that found the so-called horn structure, which was associated with the onset of deconfinement. The sharp peak in those ratios was investigated~\cite{Cleymans:2004hj} within the statistical model, which concludes that the observed behavior is due to the transition from baryon to meson production at freeze-out temperature. This behavior is maybe more important at lower energies, as has been postulated at \nica energies~\cite{Ayala:2023lnl}. 
Strange hadron production has also been analyzed using the combination model, looking for universal behavior of the anti-baryon-to-baryon yield ratios~\cite{Wang:2013duu}. In addition, the strangeness suppression factor ($\gamma_s$) has also been used to explain the horn structure in the strangeness-to-no-strangeness ratios.

Thermodynamic properties of the system created in heavy-ion collisions can be investigated through the \qgp phase diagram, drawn as a function of baryon chemical potential and temperature, by fitting statistical hadronization models~\cite{Becattini:2009sc}. Freeze-out parameters, such as temperature and baryon chemical potential, can be predicted from the multiplicity generated in heavy-ion collisions, using the statistical Grand Canonical Ensemble ~\cite{Oeschler:2006wkw}(GCE). The centrality dependence of chemical freeze-out parameters has also been investigated for \start data at an energy of 130 \gev~\cite{Cleymans:2004pp}. 

Experimental data analysis enables measurement of freeze-out parameters and determination of the location of the critical point when the system undergoes a transition phase described by the Quantum Van der Waals equation of state(\qvdw)~\cite{Vovchenko:2015vxa}. It is also possible to study the fluctuation of conserved charges in heavy ion collisions within this \qvdw model~\cite{Monnai:2019hkn,Poberezhnyuk:2018mwt}.

The present work analyzes multiplicity for identified particles from \pbpb collisions collected by \naexp experiment, \auau collisions from \ags at \sps, and \start experiment at \rhic, in an energy range from 2.7 to 200 \gev, using a statistical \thermal ~\cite{Vovchenko:2019pjl} model. This work performed a detailed analysis to get the thermodynamic properties of the system, as well as the behavior of the skewness and its relationship with the critical point in the \qvdw model.
\\
The structure of the work is as follows: section~\ref{model} describes the statistical model, as well as the methodology used in this analysis. Section~\ref{data} provides information on experimental data taken in the study. Section~\ref{results} presents the details of the results and their respective discussion, extrapolating to energies that will be covered by \nica and \fair facilities.
Finally, we draw conclusions and perspectives in section~\ref{conclusion}.

\section{Model and analysis methodology}\label{model}
Thermodynamic properties of the system created in heavy-ion collisions can be explored using thermal statistical models, assuming thermodynamic equilibrium and conservation of quantum numbers. These requirements are implemented in the \gce within the \thermal~\cite{Vovchenko:2019pjl} framework, which can be used to analyze data from heavy-ion collisions at different energies.


\subsection{\thermal}
\thermal package contains statistical thermal models: Grand Canonical Ensemble (\gce), Canonical Ensemble (\ce), and Mixed Canonical Ensemble (\mce ). This package was developed to analyze hadrons produced in relativistic heavy-ion collisions. It simulates a fireball as the starting point, similar to a sphere formed in heavy-ion collisions. Its evolution produces hadrons and their densities, which can be analyzed to characterize the system. The model allows the use of thermodynamic potentials associated with each quantum number.\\

\subsection{Grand Canonical Ensemble}
This ensemble describes a more realistic physical situation, allowing us to compute the properties of systems such as \qgp using average particle production. The model is widely used because thermodynamic potentials enable the calculation of averages of various quantities that are comparable to experimental results. The model can produce variables associated with a system considered as a Hadron Resonances Gas  (\hrg) of volume $V$ and temperature $T$ through  the logarithm of the partition function,  as follows:
\begin{equation}
 \ln Z^{GC}(T,V,\mu_{i})= \sum_{especies i} \frac{g_{i}V}{(2\pi)^{3}} \int d^{3} p \ln (1 \pm e^{- \beta(E_{i}- \mu_{i})})^{\pm1}
\end{equation}
where $ g_{i} $ correspond to the degeneracy and   $\mu_{i}$ is the chemical potential for each specie of hadron $i$, $\beta = 1/T$ and $E = \sqrt{p^{2}+m^{2}_{i}} $. Within relativistic heavy-ion collisions, the number of particles is not conserved, but rather the baryon number (\ba), strangeness (\s), and charge (\q). The chemical potential for each species of particle ($\mu_{i}$) is given by

\begin{equation}
 \mu_{i} = B_{i}\mu_{B}+S_{i}\mu_{S}+Q_{i}\mu_{Q}
\end{equation}
and the multiplicity is given by:
\begin{equation}
 N_{i}^{GC}=T\frac{\partial \ln Z^{GC}}{\partial\mu_{i}}.
\end{equation}

\noindent
Using this multiplicity distribution, we can predict particle-to-antiparticle ratios, including those particles not included as input, such as multi-strange hadrons. The baryon chemical potential, temperature, and the radii of the fireball can also be predicted; all of them will be discussed in the following sections.

\section{Data analyzed}\label{data}
To perform a study over a width energy range, we recopilate experimental data for \auau collisions at 0-5 \% from \start~\cite{STAR:2003ryp,STAR:2003jwm,STAR:2010dor,STAR:2008med,STAR:2005gfr,STAR:2017sal}, and from \ags ~\cite{Andronic:2005yp,E802:1996owm,E-802:1998xum,E866:1999ktz,E877:1999qdc,E866:2000dog,E895:2001zms}, as well as 0-6\% for \pbpb from \naexp~\cite{NA49:2002pzu,NA49:2004mrq,NA49:2006gaj,NA49:2007stj,PHENIX:2004vdg}, we know that some experiments also have \pp data; but for this analysis only results of ions were taken into account. The Table ~\ref{Tab.Multiplicity} shows the multiplicity for different hadrons, reported by the experiments at collision energy ranges from 2.7 to 200 \gev, as indicated in the first column of the Table. This will be used to study the evolution and properties of the matter created in such experiments.

 \begin{sidewaystable}
   \caption{Multiplicity results from experiments at different collision energies. These results are analyzed to get the thermodynamic properties of the system created by the experiments.}\label{Tab.Multiplicity}
\centering {\tiny
    \begin{tabular}{l l l l l l l l l}
    \hline\noalign{\smallskip}
    $\sqrt{s_{NN}}$ (GeV)& $\pi^{+}$ &  $\pi^{-}$ & $K^{+}$ &  $ K^{-}$ &  $P$ & $\bar{P}$ & $\Lambda^{0}$ & $\bar{\Lambda^{0}}$ \\ 
    \noalign{\smallskip}\hline\noalign{\smallskip}
        2.7 \cite{Andronic:2005yp,E866:2000dog,NA49:2007stj} &  14.01& 22  & 0.381  & 0.02 & 90 & & 0.4 & \\
       
        3.32 \cite{Andronic:2005yp,E866:2000dog,NA49:2007stj}& 26.4 & 26.38  & 2.34  & 0.19 & 75 & & 3.001& \\
       
        3.84 \cite{Andronic:2005yp,E866:1999ktz,E866:2000dog,NA49:2007stj}& 38.9 & 38.85  &  4.84 & 0.61 & 71 & & 5& \\
       
        4.3 \cite{Andronic:2005yp,E877:1999qdc,E866:2000dog,NA49:2007stj}& 49.7 & 49.6  & 7.85  & 1.26 & 67 & & 7 & \\
       
        4.86 \cite{Andronic:2005yp,E802:1996owm,E-802:1998xum,E802:1999hit,E866:2000dog,NA49:2007stj}& 57.1 & 71  & 11.55  & 2.21 & 60 & & 10 &\\
        
        6.3 \cite{Andronic:2005yp,NA49:2006gaj,NA49:2007stj}& 72.9 &  84.8 &  16.4 & 5.58 & 46.1 & 0.06 & 11& \\
       
        7.6 \cite{Andronic:2005yp,NA49:2006gaj,NA49:2007stj}& 83 & 96.5  & 21.2  & 7.8 & 42.1 & 0.16 & 10.4& 0.15 \\
       
        7.7 \cite{STAR:2017sal} & 93.4 & 100  & 20.8  & 7.7 & 54.9 & 0.39 & & \\
        
        8.7 \cite{Andronic:2005yp,NA49:2002pzu,NA49:2006gaj,PHENIX:2004vdg}& 96.6 &  106.1 & 20.1  & 7.58 & 41.3 & 0.32 & 10 & 0.9 \\
        
        11.5 \cite{STAR:2017sal} & 123.9 & 129.8  & 25  & 12.3 & 44 & 1.5 & & \\
       
        12.3 \cite{Andronic:2005yp,NA49:2002pzu,NA49:2006gaj,PHENIX:2004vdg}& 132 &  140.41 & 24.6  & 11.7 & 3.1 & 0.87 & 9 & 1.4\\
       
        17.3 \cite{Andronic:2005yp,NA49:2002pzu,NA49:2006gaj,NA49:2007stj,PHENIX:2004vdg}&  170.1&  175.4 & 29.6  & 16.8 & 29.6 & 1.66 & 9.3& 9\\
       
        19.6 \cite{STAR:2017sal}& 161.3 &  165.8 &  29.6 & 18.8 & 34.2 & 4.2 & & \\
        
        27 \cite{STAR:2017sal}&  172.9 & 177.1  &  31.1 & 22.6 & 31.7 & 6& & \\
        39 \cite{STAR:2017sal} & 182.3 &  185.8 & 32  & 25 & 26.5 & 26.5& & \\
        62.4 \cite{STAR:2008med}& 237 & 233  & 37.6  & 32.4 & 29 & 13.6& & \\
        130 \cite{STAR:2008med}& 278 &  280 &  46.3 & 42.7 & 28.2 & 20& & \\
        200 \cite{STAR:2008med}& 322 &  327 &  51.3 & 49.5 & 34.7 & 26.7 & & \\         \noalign{\smallskip}\hline
    \end{tabular}
   }  
\end{sidewaystable}

\subsection{Yields for the analysis}
The yield from \start data correspond to light hadrons: $\pi^{\pm}$, $K^{\pm}$, $p$ y $\bar{p}$), as well as  strangeness baryons: $\Lambda$, $\bar{\Lambda}$ and $\Xi$, produced in \auau collisions, over an energy range of $\sqrt{s_{NN}}=7.7 -200$ \gev, and rapidity $|y| < 0.1$.
The primary pions include feed-down corrections from weak decays, muon contamination, and background pions produced in the detector materials. The pions' daughters from the weak decays of $K_s^0$ and $\Lambda^0$ are identified in simulations. The total pion background contribution from weak decay decreases with increasing $p_{\rm T}$, a contribution that was estimated for energies $\sqrt{s_{NN}}= 7.7-39 GeV$.
The proton yields do not include feed-down corrections, but a cut on the distance of closest approach ($\rm{DCA}$) between each track and the event vertex was applied ($\rm{DCA}<3$ cm), reducing a small fraction of protons that come from hyperon decays. Therefore, the proton and anti-proton yields presented here are inclusive.
The systematic uncertainty is estimated by varying the vertex along the $z$ direction. The track cuts are also varied across the $\rm{DCA}$, and the number of sigmas cut on the energy loss is used to obtain a prediction of $m^2$ distributions for the purity of the hadrons.
In general, charged baryons are reported as inclusive to avoid additional systematic uncertainties. More details on the yields measured can be found in Ref.~\cite{STAR:2017sal} and references therein.

Experiment $\rm{E802}$ at \ags measured yields from \auau collision  in energy range of $\sqrt{s_{NN}} =2.7 - 4.86$ GeV, with rapidity,   $|y| < 0.4$~\cite{E802:1996owm}. The \naexp made \pbpb collisions in the energy range $\sqrt{s_{NN}}= 6.3-17.3 $ GeV, and in rapidity $|y| < 0.6$.
The measurements use different selection criteria; for instance, pions and protons are considered inclusive.
The yields from \naexp~\cite{NA49:2002pzu,NA49:2006gaj,NA49:2004mrq} and $\rm{E802}$ at \ags have higher uncertainties than those from \start, fewer statistics, and results for a few energies, so it is not possible to make a quantitative comparison with \start data. A review of different data sets at different energies and from different experiments has been reported~\cite{Fischer:2022oqp}. 
Consequently, our analysis focuses on \start data, and some fits are extrapolated to lower energy for qualitative comparison with \ags and \naexp data.

\section{Results and discussions}\label{results}
Using \thermal packages and data of Table~\ref{Tab.Multiplicity}, we analyze the results described in the following subsections.

\subsection{ Multiplicity and ratios} 
\begin{figure}[h!]
  \centering
  \includegraphics[width=1.00\linewidth]{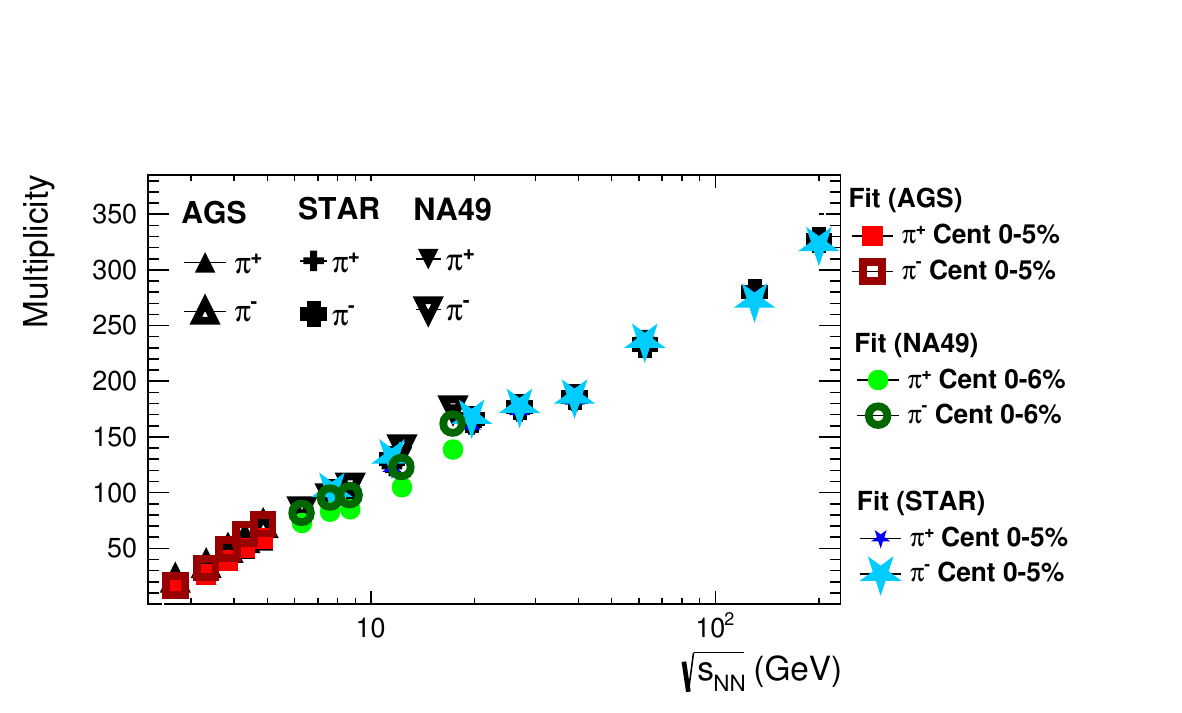}
    \includegraphics[width=1.0\linewidth]{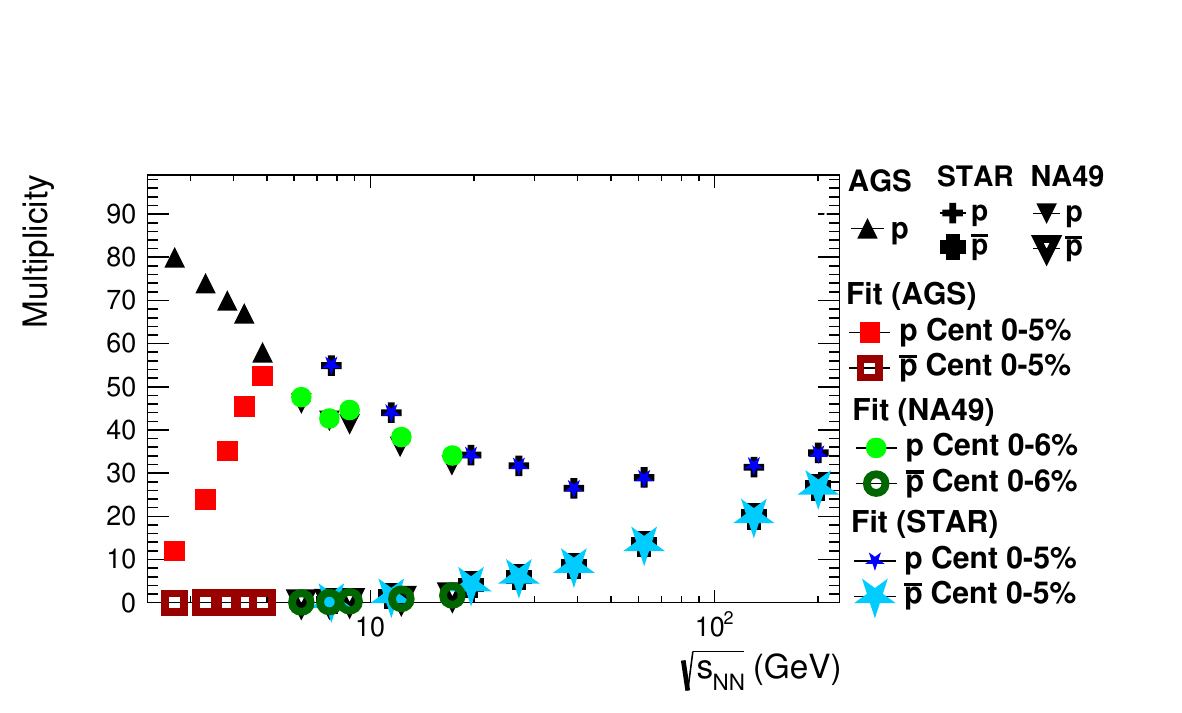}
    \caption{Multiplicity of $\pi^{\pm}$ (top) and  $p, \bar{p}$ (bottom). Comparisons between experimental results and those predicted from the model are also shown.}\label{Fig.Multiplicities1}
\end{figure}

\begin{figure}[h]
  \centering
            \includegraphics[scale =0.45]{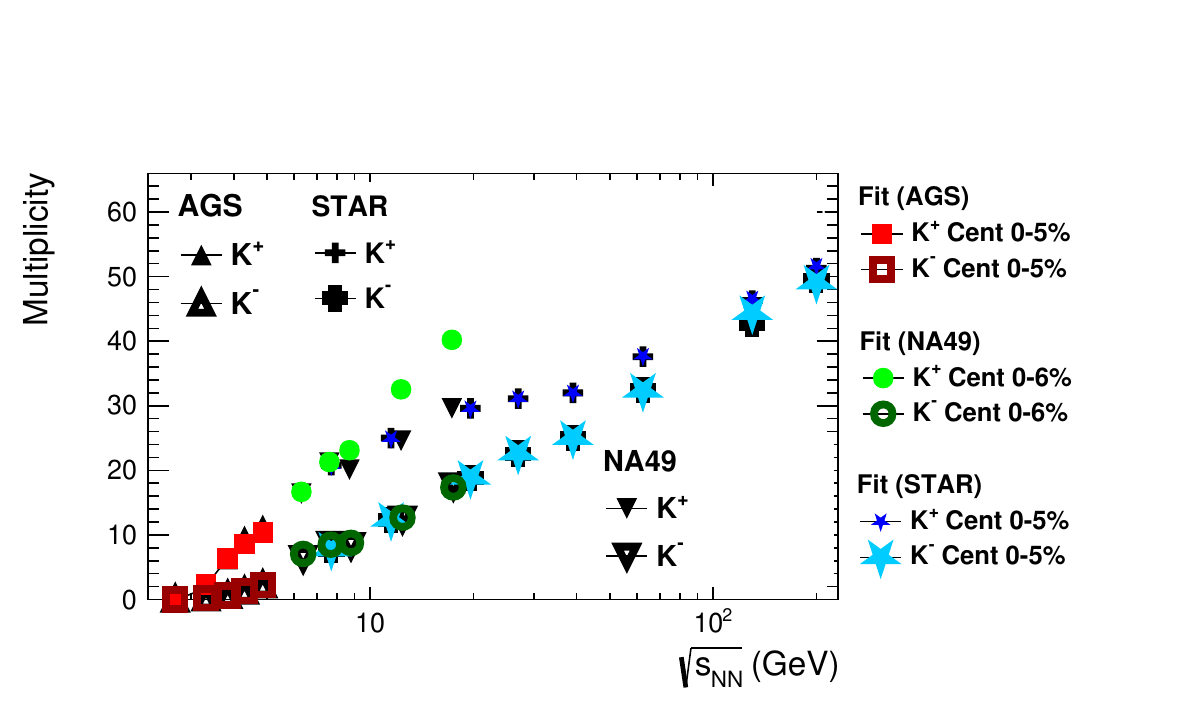}
        \includegraphics[scale =0.45]{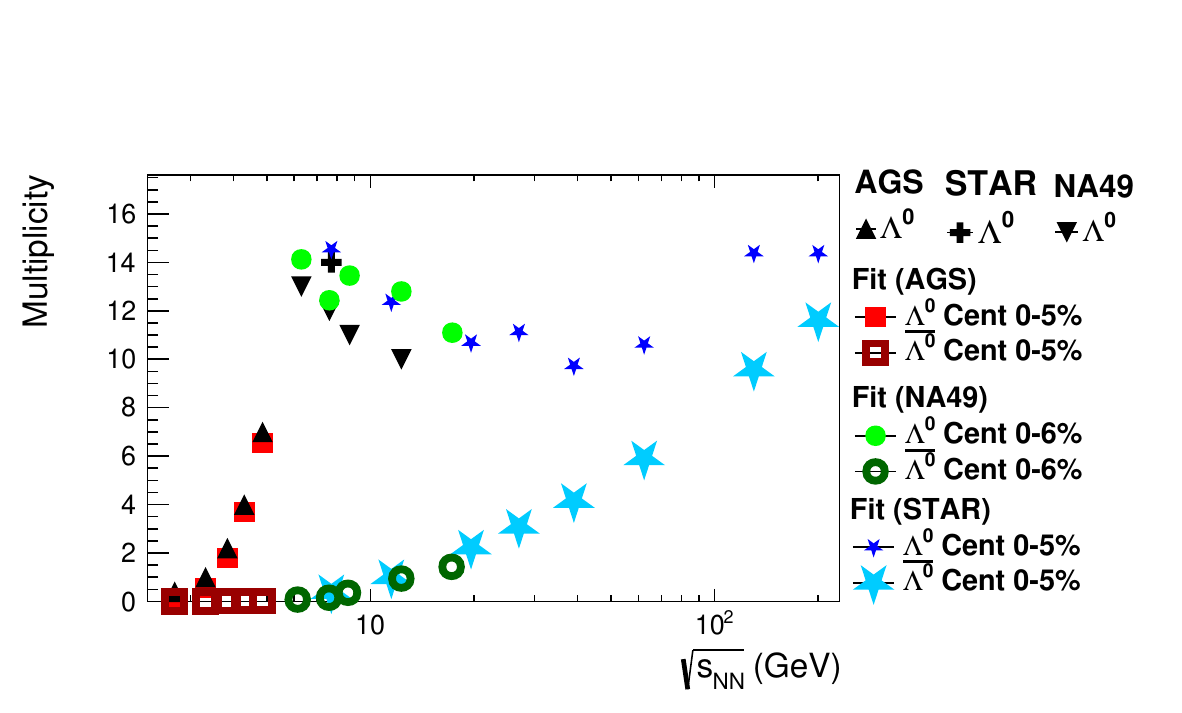}
    \caption{Multiplicity of  $K^{\pm}$ (top)  and $\Lambda, \bar{\Lambda}$  (bottom). Comparisons between experimental results and those predicted from the model are also shown.}\label{Fig.Multiplicities2}
\end{figure}

Considering the known multiplicity as an input parameter, one can predict the multiplicity for unknown hadrons. For instance, the Table~\ref{Tab.Multiplicity} shows no measurements of  $\bar \Lambda$ for almost all energies; however, using the measured particles, we can predict multiplicities for many strangeness hadrons, such as $\Xi,  \Omega, \Sigma$, and $\phi$. The Fig.~\ref{Fig.Multiplicities1} shows the multiplicity computed and compared to experimental measurements for pions and protons for a width range of energy measured by experiments \ags and \start in centrality 0-5\%, and  \naexp in 0-6\%. The pion case has almost exponential growth with the collision energy. The protons measured by \ags show a decrease with the energy, while the protons predicted by the model increase for the energy range of 2.7-4.86 GeV. This behavior occurs because, at collision energies below 6.3 GeV, the protons that are used as input are not measured in the experiment; therefore, the prediction worsens. Increasing the collision energy, the proton multiplicity decreases, while the anti-proton multiplicity increases. These results are completely understandable because at very low energy, it is not possible to produce antiprotons. 
Figure~\ref{Fig.Multiplicities2} shows $K^{\pm}$ (top) and $\Lambda^{0}$ (bottom), in both cases the particle production increases faster than the anti-particle, reaching a maximum around energy 6.3-7.7 GeV and 17.3 GeV for $\Lambda's$ and $K^{+}$ respectively, on the contrary, the anti-particle grows monotonously.

\begin{figure}[h]
    \centering
    \includegraphics[scale=0.45]{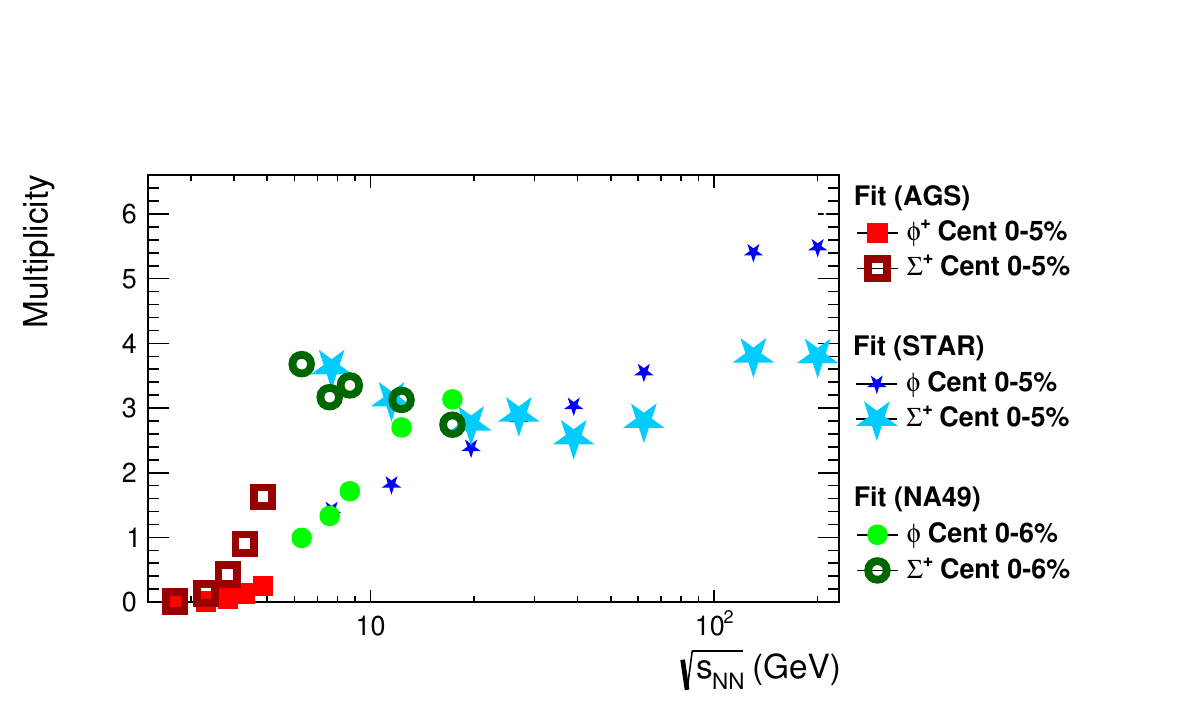}
    \includegraphics[scale=0.45]{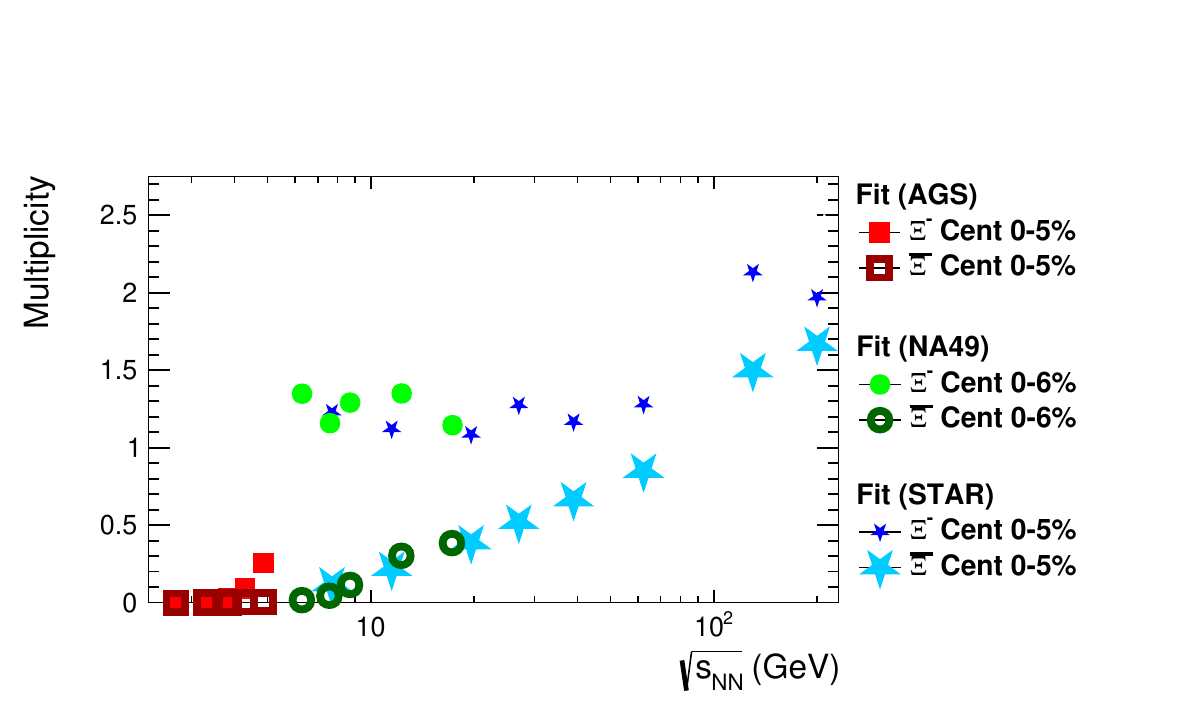}
    \includegraphics[scale=0.5]{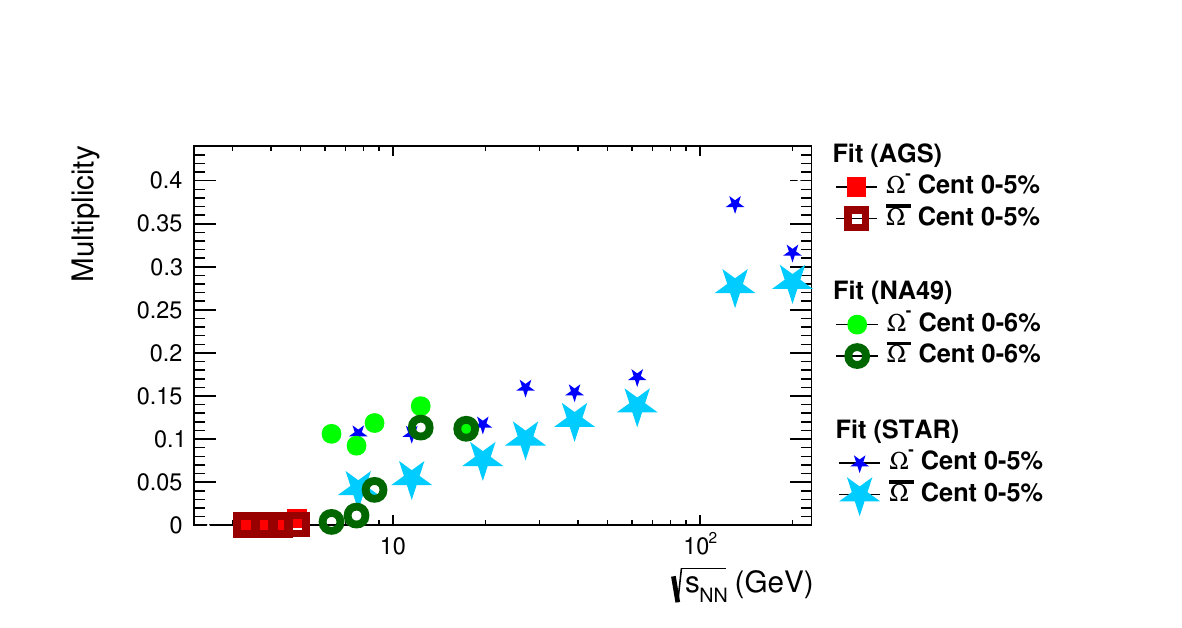}
    \caption{Multiplicity distribution for $\Omega^{\pm}$,  $\Xi$, $\bar{\Xi}$, $\Sigma^{+}$, $\phi$ }\label{Fig.MultiStrange}
\end{figure}

Data for $K^{\pm}$,  $\Lambda$, and  $\bar{\Lambda}$ are well reproduced between 5-11 GeV, with a maximum around  7.7 GeV, except for the case of \naexp results, which by the way, correspond to \pbpb collisions while the \start experiment made \auau ions.\\
Once we have validated results for the known multiplicity, the multi-strange hadrons have also been computed and shown in Fig.~\ref{Fig.MultiStrange}, for $\phi (s\bar{s}$) and $\Sigma^{+}(uus)$ (top), $\Xi^{\pm}$ (middle) and $\Omega$ (bottom). The behavior is similar to that in previous cases: the multiplicity of particles increases faster than that of antiparticles, reaching a maximum and then decreasing, whereas antiparticles exhibit monotonic growth as energy increases. Multi-strange hadrons also show a rise in multiplicity, reaching a maximum below 10 \gev, then go down, and increase. Note that the slope of the rise is faster for lighter particles.

\begin{figure}[!htb]
    \centering
    \includegraphics[scale =0.4]{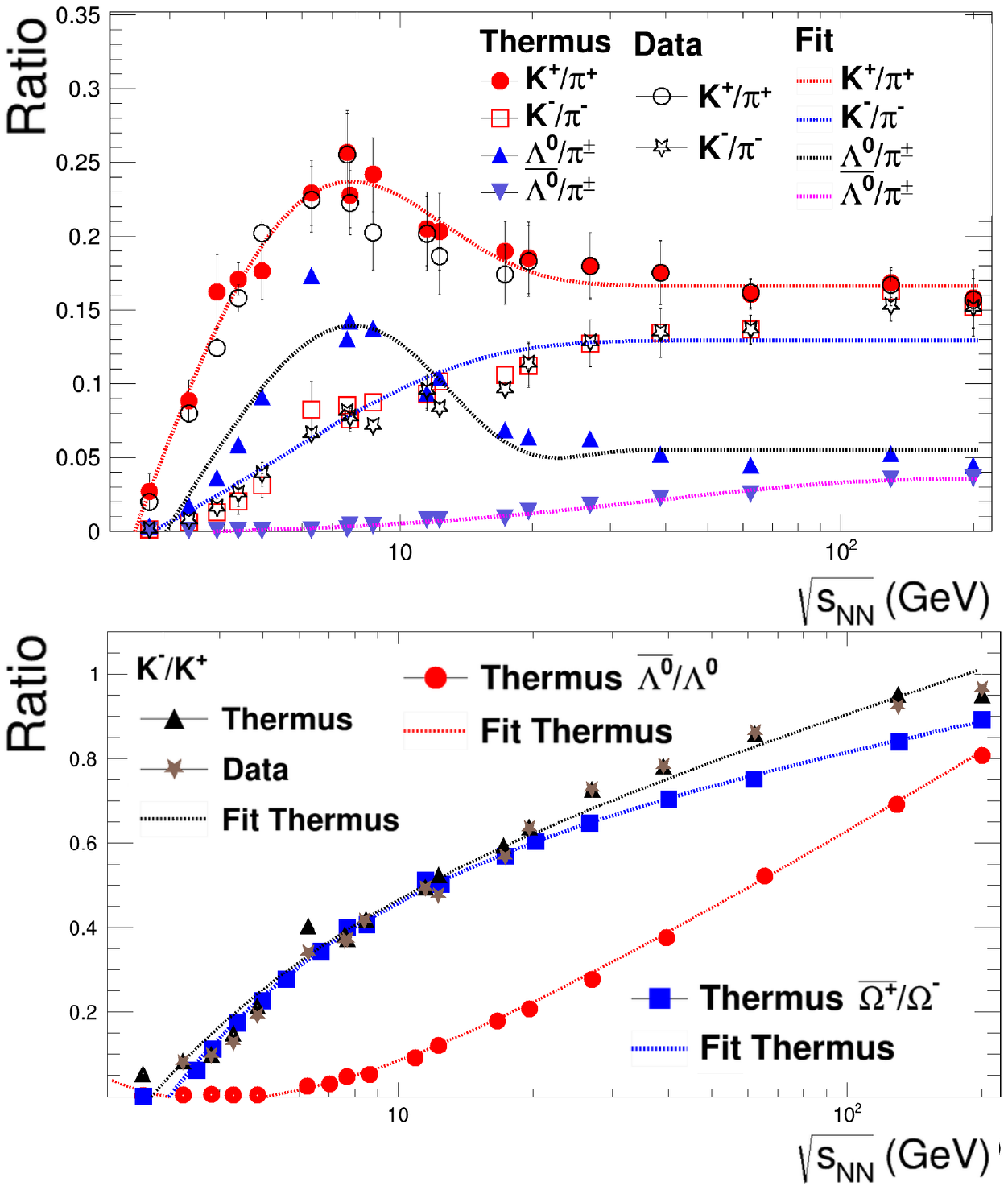}
  \caption{$K$ to $\pi$ and $\Lambda$ to $\pi$ ratios (top) and  comparison meson to anti-mesons and baryon anti-baryon ratios (bottom) }\label{LambdaToPion}   
\end{figure}

 The results of  Fig.~\ref{Fig.MultiStrange} are predictions, since they were not used as input in the calculation. The multiplicity distribution allows us to compute the meson-to-meson and baryon-to-meson ratios, as shown in  Fig.~\ref{LambdaToPion} for \KaonToPionPlus, \KaonToPionMinos, \LambdaToPi, \ALambdaToPi (top), \KaonToKaonPM, \OmegaToOmegaPM, and \ALambdaLambda (bottom). Each distribution is well described by the para\-metri\-zation given by Eq.~\Ref{Eq.ra1} with parameters listed in Table~\Ref{Tab.Eq.ra1}. Experimental results of  \KaonToPionPlus and \LambdaToPi are well reproduced by the model, providing confidence to compute ratios of particles not measured yet at those energies, such as \OmegaToOmegaPM and \ALambdaLambda, which are shown in the bottom panel of Fig.~\ref{LambdaToPion}.

  \begin{equation}\label{Eq.ra1}
Ratio=ae^{b\sqrt{s_{NN}}} \sin({c\sqrt{s_{NN}} + d}) +e
\end{equation}

  \begin{table}
     \caption{Parameters used in Eq.~\Ref{Eq.ra1}}
    \label{Tab.Eq.ra1}
    \begin{tabular}{llllll} 
      \hline\noalign{\smallskip}
      & a     & b & c & d & e\\ 
      \noalign{\smallskip}\hline\noalign{\smallskip}
        $K^+ / \pi^+$ &1.54 & 2.89&-1.24 &3.19&1.66 \\
        $K^- / \pi^-$& 2.801&1.87 &-3.76&4.02&1.29 \\
      $\Lambda^0 / \pi^\pm$  & 0.50& -0.18 & -0.21&3.95&0.05\\ 
       $\bar{\Lambda^0} / \pi^\pm$ & 0.08 & -0.02 & 0.002 &3.62 &0.03\\ 
\noalign{\smallskip}\hline
    \end{tabular}
  
\end{table}

 \subsection{Baryon chemical potential and temperature}
Thermodynamic potentials of a system play a significant role in determining its properties; some of them are discussed in the rest of this section. Baryon chemical potential predicted from the experimental data can be parametrized as a function of the collision energy~\cite{Randrup:2006nr}:
\begin{equation}\label{Eq.Mu}
\mu_{B}(\sqrt{s_{NN}}) = \frac{d}{1+e\sqrt{s_{NN}}}
\end{equation}
where  $d=1.308\pm 0.028$ \gev,  $e=0.272\pm 0.008$\gev$^{-1}$ and the temperature  is parametrized  by:
\begin{equation}\label{Eq.T}
T(\sqrt{\mu_{B}}) = a-b\mu_{B}^{2}-c\mu_{B}^4
\end{equation}
with $a=0.166\pm 0.002$ GeV, $b=0.139\pm 0.016$ GeV, and $c=0.053\pm 0.021$ \gev. Equations ~\ref{Eq.Mu} and ~\ref{Eq.T} describe with high precision the \start data~~\cite{STAR:2003ryp,STAR:2003jwm,STAR:2010dor,STAR:2008med,STAR:2005gfr,STAR:2017sal}. 

\begin{figure}[h]
    \centering
    \includegraphics[width=1.3\linewidth]{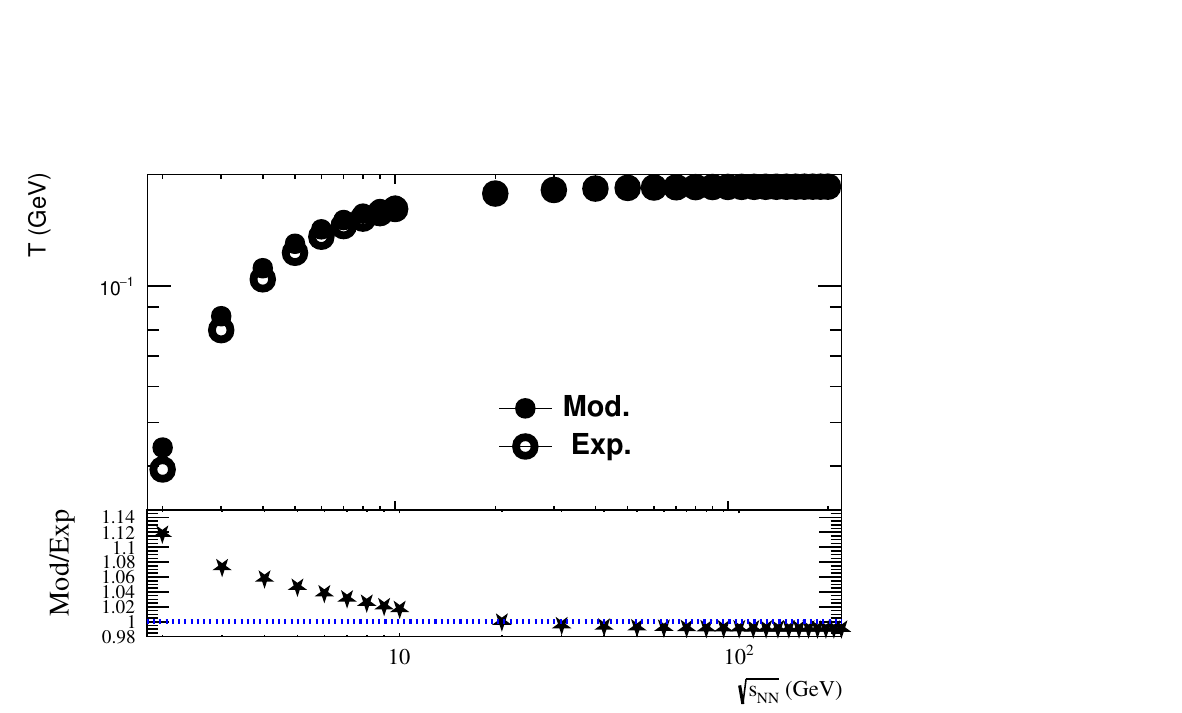}
    \caption{Comparison between the temperature reported by \start and a redefined parametrization. A discrepancy of up to 13\% is observed at lower energies.} \label{Fig.muvsT}
\end{figure}

However, this parametric function of the temperature deviates from the data reported at lower collision energies, so we re-parametrized the expression using data collected from 2.7 to 200 \gev, with Eq.~\ref{Eq.NMu}.

\begin{equation}\label{Eq.NMu}
T(\sqrt{\mu_{B}}) = a_{T}-b_{T}\mu_{B}^{2}-c_{T}\mu_{B}^4+f_{T}\mu_{B}^6
\end{equation}
The new parameters predicted from the data are:  $a_{T}=0.16446 \pm 0.0012$ \gev, $b_{T}=0.11196 \pm 0.014$ \gev, $c_{T}=0.139139 \pm 0.032$ \gev, $f_{T}=0.04637 \pm 0.0012$,   $d=1.27347 \pm 0.0023$ \gev  and $e=0.28554 \pm 0.00012$ \gev, the last two correspond to Eq.~\ref{Eq.Mu}.
With these new parameters, we observe an improvement in the data description, as shown in Fig.~\ref{Fig.muvsT}. The lower panel of the figure shows the ratio of the model to the data, where we observe good reproduction of the data at energies above 10 \gev, with a discrepancy of up to 13\% at lower collision energies. The apparently insignificant difference is crucial for upcoming experiments such as \mpdnica~\cite{MPD:2025jzd}, where detailed temperature analysis is essential at low collision energy.

\begin{figure}[h]
    \centering
    \includegraphics[width=1.3\linewidth]{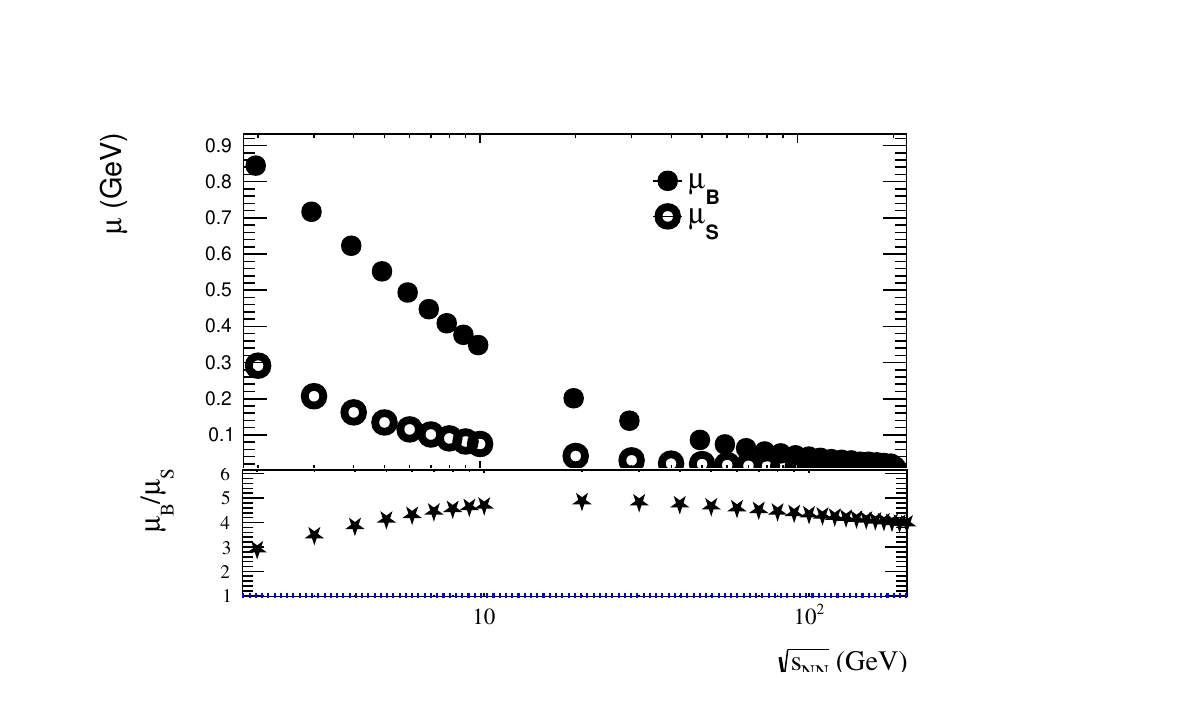}
    \caption{ Baryon chemical and strangeness potentials as a function of collision energy.}\label{Fig.Potentials}
\end{figure}

The strangeness chemical potential has also been para\-metri\-zed as a function of collision energy by Eq.~\ref{Eq.Mu}, with parameters $d =1.075\pm 0.079$ and $e= 1.366\pm0.123$.  The baryon chemical potential increases as the energy decreases, with a similar behavior for the strangeness chemical potential, but it increases more slowly, see Fig.~\ref{Fig.Potentials}. The bottom panel of the figure shows the baryon chemical potential to strangeness chemical potential ratio; it is evident that the baryon chemical potential is more than twice the strangeness chemical potential at lower energies, but the ratio can reach 5 at energies around 20 \gev.\\

\begin{figure}[htb]
    \centering
    \includegraphics[scale=0.19]{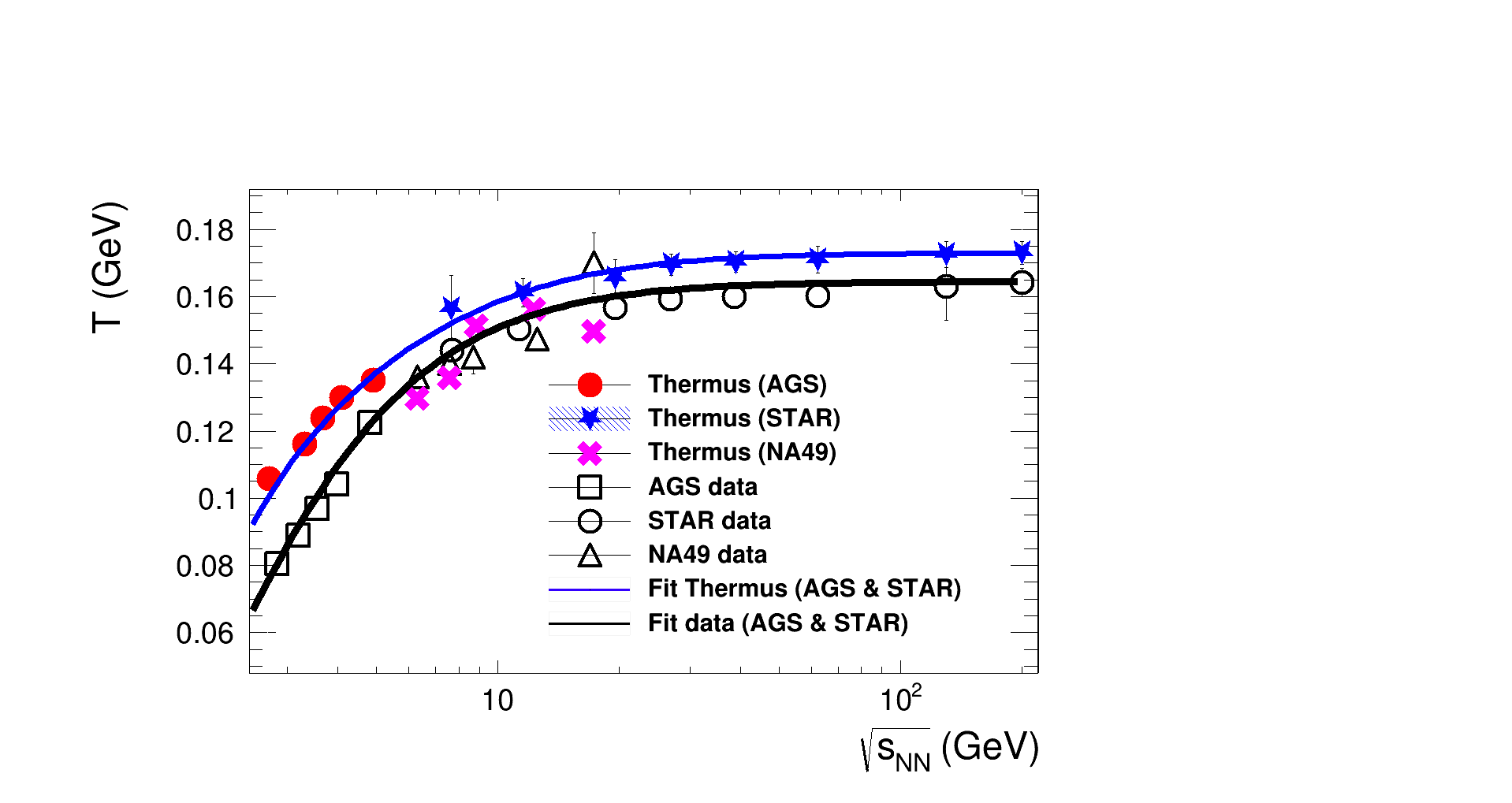}
    \caption{Temperature of the model (fully markers) and experimental measurements (empty markers), as well as fits (curves).}\label{Fig.Temperature}  
\end{figure}

\begin{figure}[htb!]
    \centering
     \includegraphics[scale=0.55]{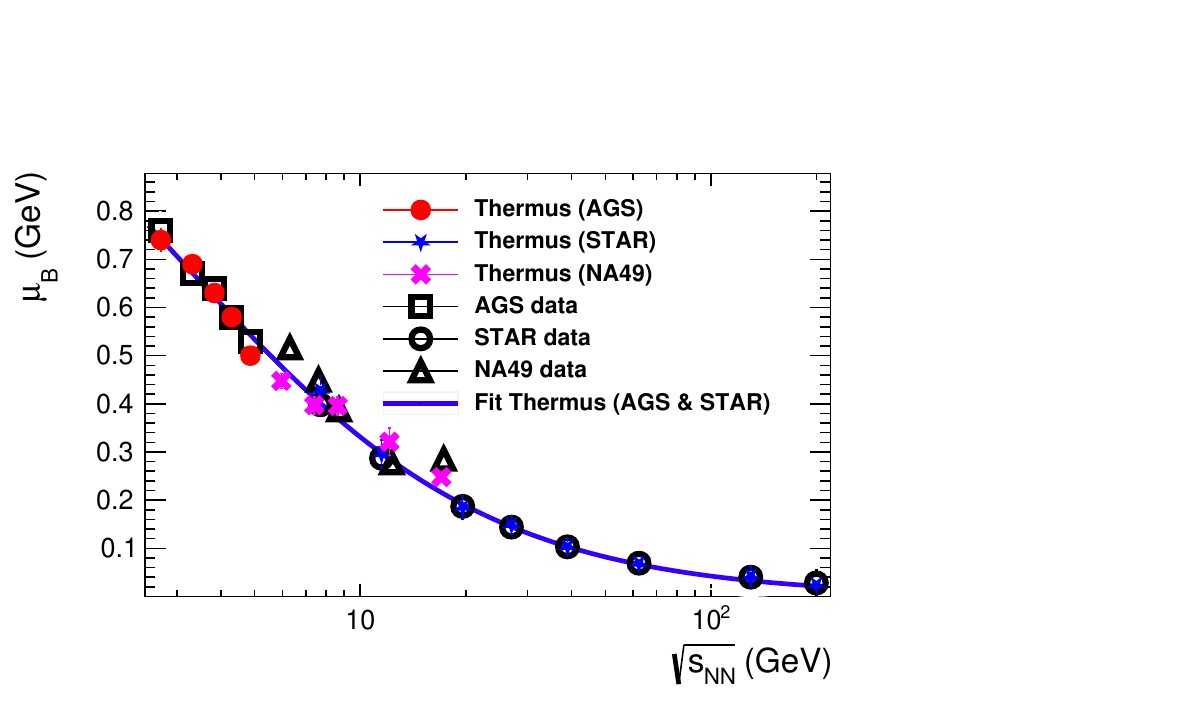}
    \includegraphics[scale=0.55]{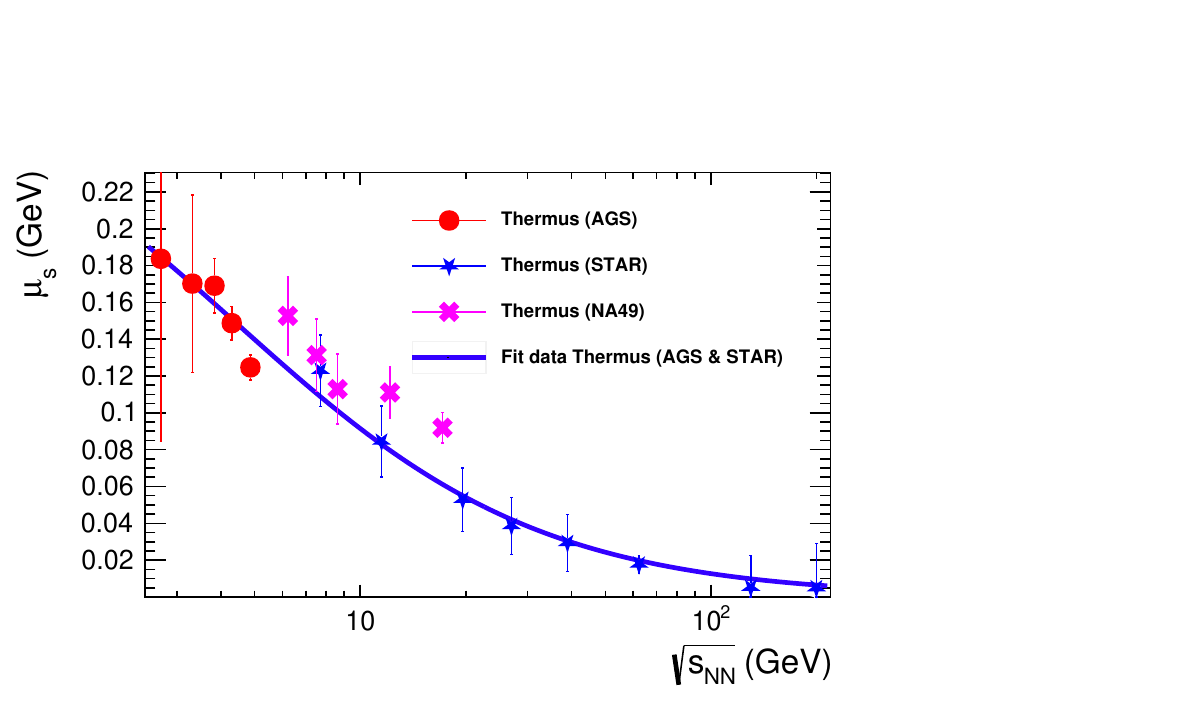}
    \caption{Baryon Chemical and strangeness potential predicted from experimental multiplicities at different energies.} \label{Fig.BSPotential}
\end{figure}

Previous multiplicity results enable us to predict the temperature as a function of the baryon chemical potential and collision energy. The temperature is shown in Fig.~\ref{Fig.Temperature},  where the results of the model (indicated by full markers) and the data reported from different experiments (empty markers) are compared and fitted (curves).
For higher energy, the temperature reaches values of $~164$ \mev and $~173$ \mev, with clear windows between those reported in the literature and our results.  It is worth noting that at lower collision energy, the multiplicity is lower; consequently, the errors are larger than at higher energy. A clear example of this is Fig.~\ref{Fig.BSPotential}, which shows that results from \naexp and \ags have larger error bars compared to \start data.
Another source of this temperature discrepancy is the radii used in the \thermal model; we get $T_{mod}$) with free parameters, while for \start, $T_{data}$  was obtained by fixing  $\mu_{s}$ and  $\gamma_{s}$.  
Using Eq.~\ref{Eq.NMu} we fit the temperature, the parameters of the fit are shown in Table~\ref{tab.ParaTemperature}.

\begin{table}[htb!]
\caption{Parameters predicted by fitting Fig. ~\ref{Fig.Temperature} of temperature with Eq.~\ref{Eq.NMu}} 
    \label{tab.ParaTemperature}  
    \begin{tabular}{llllll} 
      \hline\noalign{\smallskip}
        T(GeV)       & $a_T$ & $b_T$ & $c_T$ & $f_T$ & $\chi^2/ndf$\\  
        \noalign{\smallskip}\hline\noalign{\smallskip}
        $T_{data}$ & 0.164  &  0.111 & 0.139  & 0.046 & 0.5421 \\
           &  $\pm$ 0.001 & $\pm$ 0.014 &$\pm$ 0.032 &$\pm$0.001\\
        $T_{mod}$ & 0.173 & 0.143 & -0.107  & -0.157 & 0.8421\\
           & $\pm $0.002 & $\pm$ 0.04 & $\pm$ 0.02  &$\pm$ 0.02\\
  \noalign{\smallskip}\hline   
    \end{tabular}    
\end{table}

The same analysis was performed to predict the baryon chemical potential; the results agree with those reported in the literature and are shown in the top panel of Fig.~\ref{Fig.BSPotential}. Using this information, it is possible to predict the strangeness chemical potential plotted in the bottom panel of Fig.~\ref{Fig.BSPotential}. These potentials were parametrized according to Eq.~\ref{Eq.Mu}, with parameters given in Table~\ref{Tab.Potentials2}. It is worth mentioning that both potentials computed with \naexp are slightly higher with respect to the \start; this could be because the results were from \pbpb collisions instead of \auau, as the \start and \ags data. 
Continuous lines also represent the fits for both the baryon and strangeness potentials.

\begin{table}[htb!]
  \caption{Parameters for the baryon chemical and strangeness potentials}\label{Tab.Potentials2}
  \begin{tabular}{lll} 
    \hline\noalign{\smallskip}
    $\mu (GeV)$     & a     & b\\   
    \noalign{\smallskip}\hline\noalign{\smallskip}
        $\mu_{B}$ & 1.373 $\pm$ 0.086& 0.3148 $\pm$ 0.034\\
    $\mu_{s}$ & 0.295 $\pm$ 0.024 & 0.223 $\pm$ 0.036\\ 
    \noalign{\smallskip}\hline
    \end{tabular}    
\end{table}

\subsection{Volume}
\begin{figure}[h]
    \centering
        \includegraphics[scale =0.55]{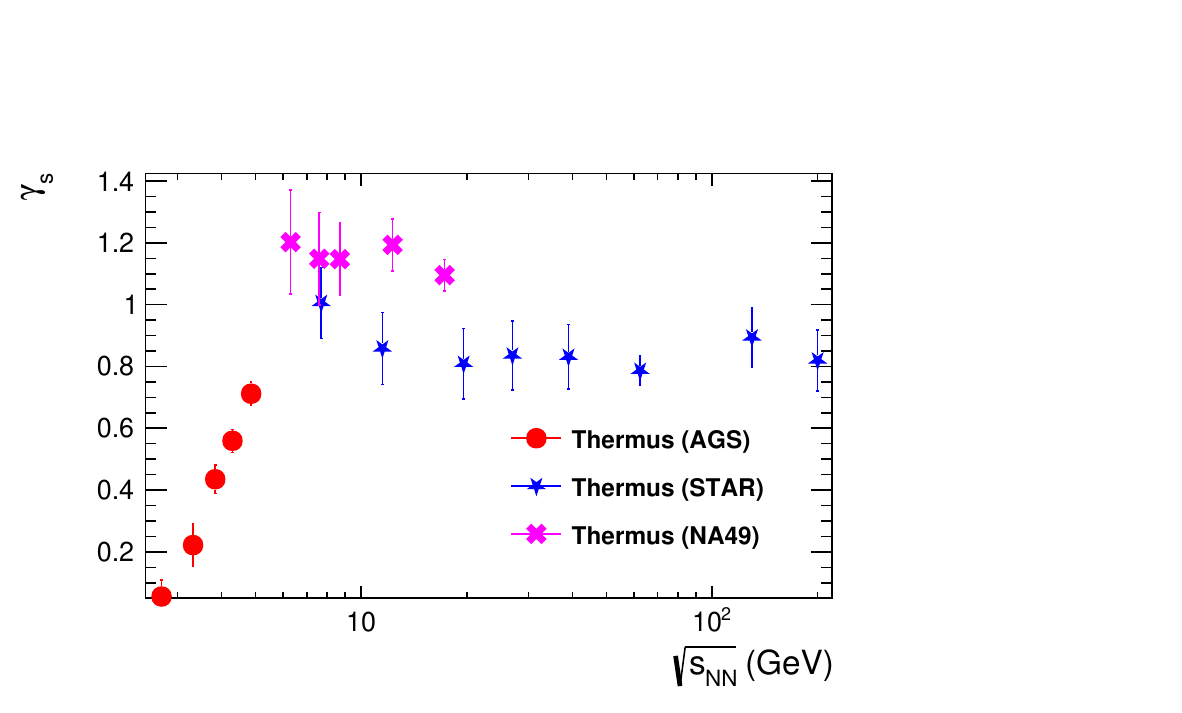}
        \includegraphics[scale =0.55]{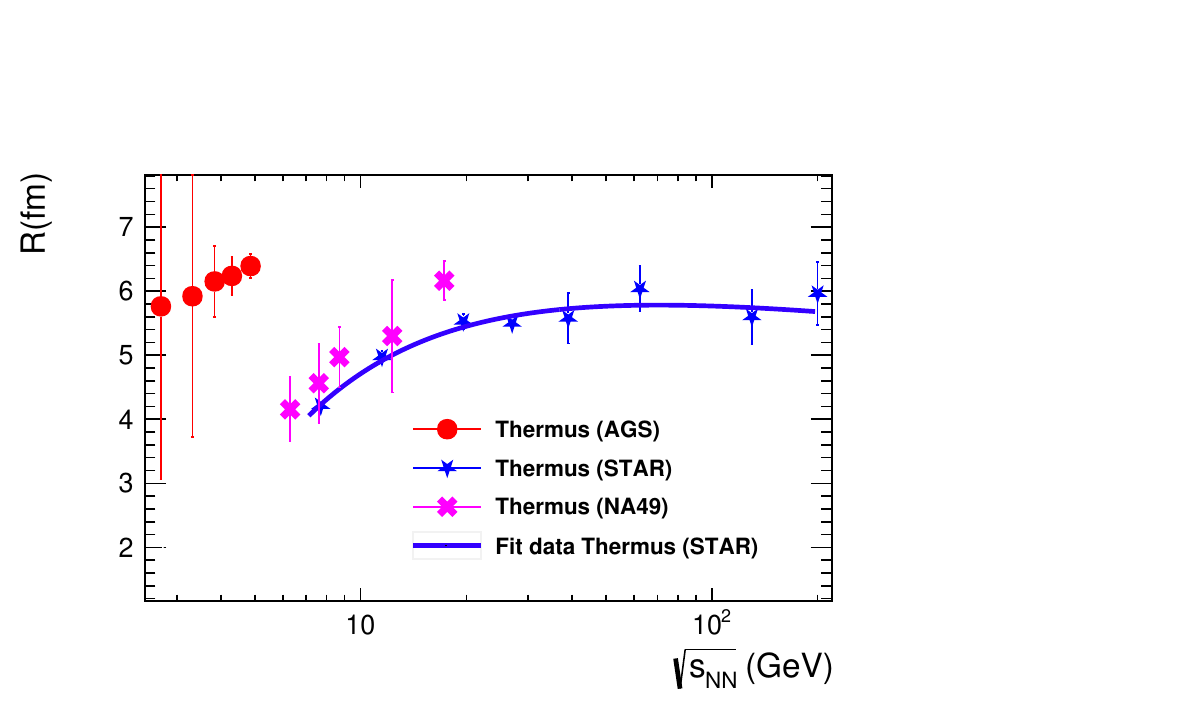}
        \caption{Strangeness suppression factor (top) and fireball radius (bottom) as a function of collision energy. The curve in the bottom panel is a fit to the STAR data.}\label{Fig.RadioGamma}
\end{figure}

In most realistic situations, the number of nucleons in the collision may fluctuate. Consequently, the system size, as determined by the model, varies with the collision energy. For the data analyzed, which are collected in Table~\ref{Tab.Multiplicity}, the radii of the fireball were also predicted, along with the strangeness suppression factor ($\gamma_s$). Those distributions are shown in Fig.~\ref{Fig.RadioGamma}, 
where the top panel shows the $\gamma_{s}$ factor, its growth reaching a maximum of around 7.7 \gev, then slightly goes down and saturates. Since the energy from \ags and \start is complementary, the distribution's maximum lies in the energy range where data from \naexp are available. However, it will take care, since \naexp has low statistics and large uncertainties. It is worth noting that \ags accounted for all protons, not only primaries, unlike \naexp and \start. The behavior of $\gamma_{s}$ indicates that the system reaches the equilibrium as the energy increases. The bottom panel shows the collision energy dependence of the radii predicted by different experiments, with the curve corresponding to the \start data fit, which saturates at around 6 $fm$ without uncertainties. Lightly higher values are observed for \naexp with respect to \start data, due to the first corresponding to \pbpb while the second to \auau collisions. The low multiplicity of \naexp also produces a large uncertainty. The highest values of \ags are observed due to insufficient species and their low multiplicity, which limit the precision of the radii.
Consequently, the model produces significantly large uncertainties, attempting to compensate for the lack of particles by increasing the fireball's radius.\\

\section{Analysis by centrality classes}
\begin{figure}[h!]
    \centering
    \includegraphics[scale=0.6]{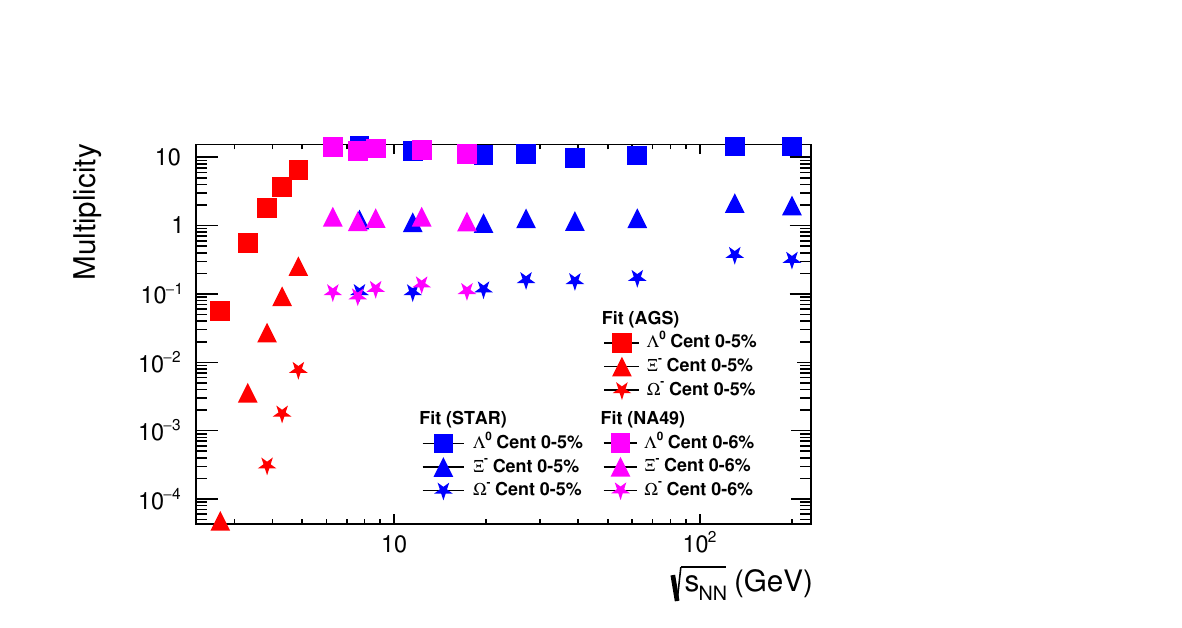}
        \includegraphics[scale=0.6]{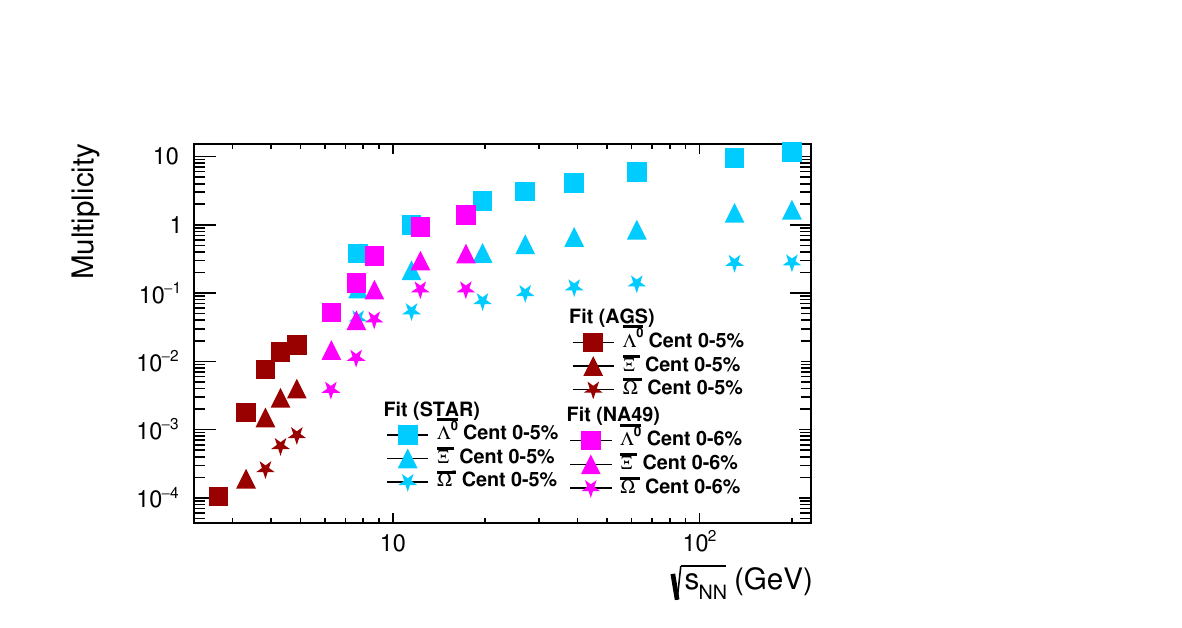}
    \caption{Multiplicity distributions for multi-strange hadrons: 1-strange ($\Lambda$), 2-strange ($\Xi$), and 3-strange ($\Omega$) for particle (top) and anti-particle (bottom)}\label{Fig.MultByCentrality}
\end{figure}

In heavy-ion collisions, the centrality variable allows us to follow the evolution of the system created by comparing results from peripheral to central collisions, such as hadron production. At the most central collisions, the production is higher for hadrons with less content of strangeness, as can be seen in the  Fig.~\ref{Fig.MultByCentrality},
where it's compared multiplicity for hadrons with 1-strange ($\Lambda^0(uds)$), 2-strange ($\Xi^-(dss)$), and 3-strange ($\Omega^-(sss)$). Starting from heavy 3-strange, it scales for the lightest baryons. Hadrons exhibit a steep growth (top panel) and saturate for collision energies exceeding 10 \gev. In contrast, anti-hadrons grow monotonically with the collision energy, as shown in the bottom panel of the figure. Our analysis predicts strangeness production at very low energies, mainly for $\Omega$ and $\Xi$, with a behavior consistent with the measured data.

\begin{figure}[h!]
    \centering
    \includegraphics[scale =0.45]{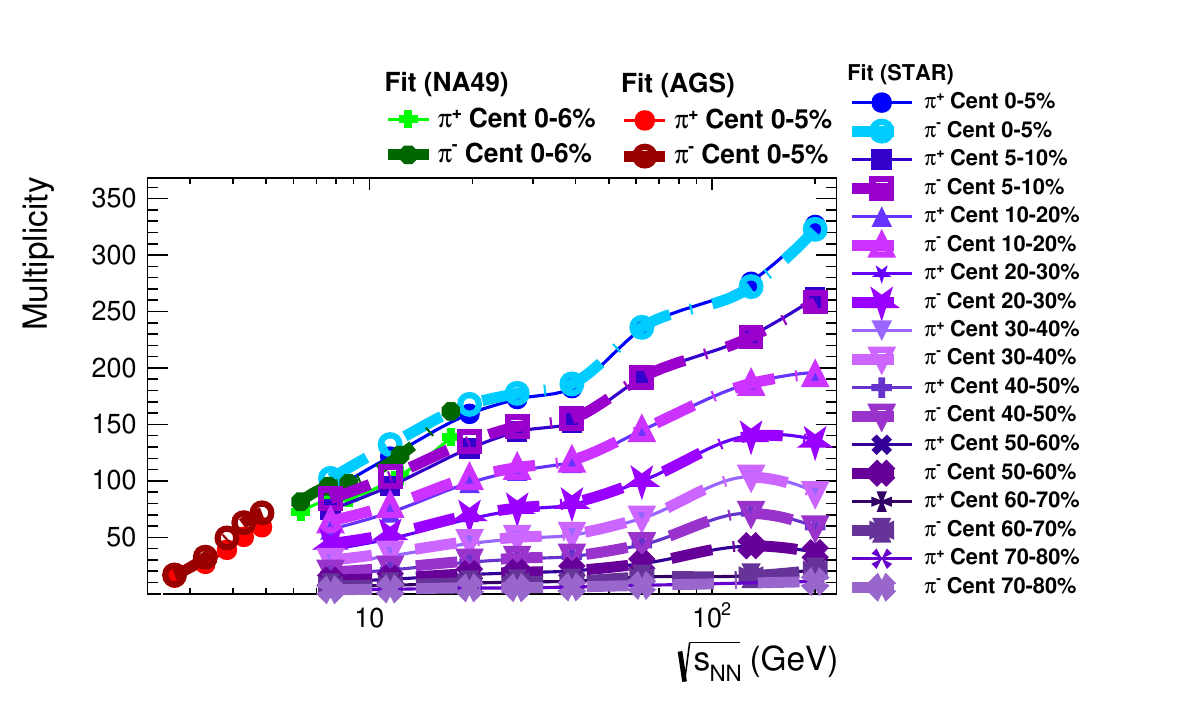}
    \includegraphics[scale =0.45]{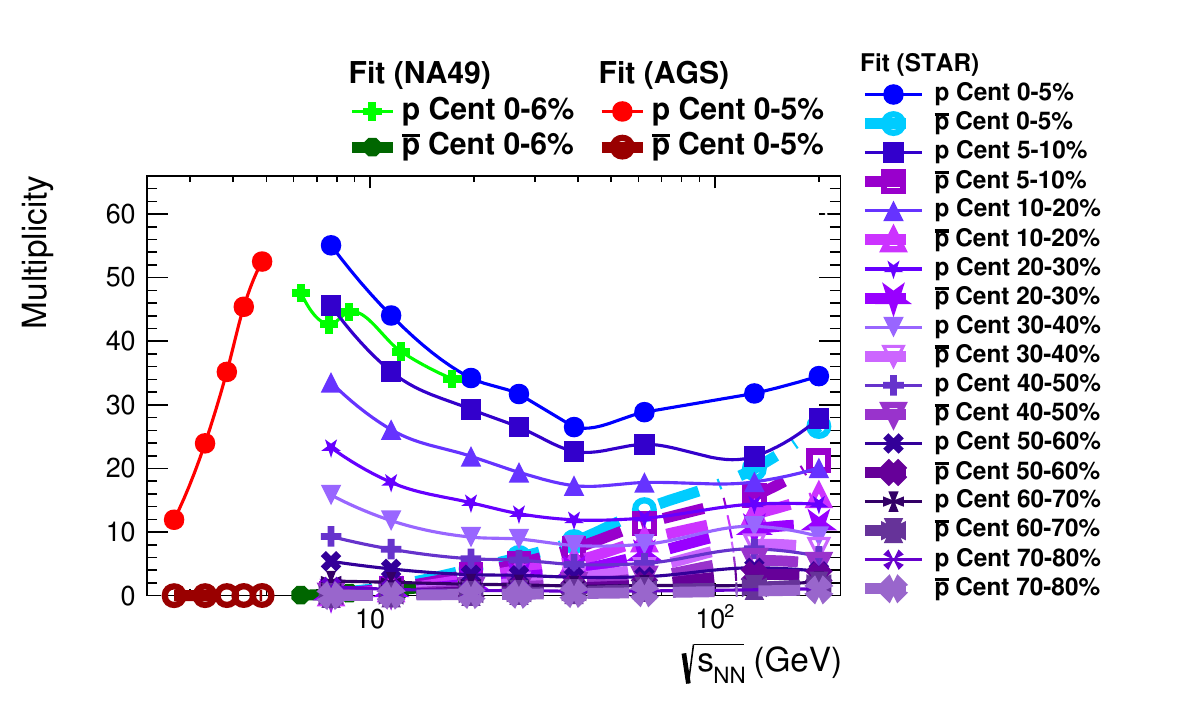}
    \caption{Multiplicity for $\pi^{\pm}$ (top) y $p, \bar{p}$ (bottom) as a function of collision energy and for centrality bins.
    }\label{Fig.Nch.Cen.pi.pro}
\end{figure}

Analyzing the centrality classes of the \start data, we can also predict 
multiplicities for each hadron species; for instance, the Fig.~\ref{Fig.Nch.Cen.pi.pro} shows the energy dependence of the multiplicity for $\pi^{\pm}$ (top) and for $p$ and $\bar{p}$ (bottom), for nine centrality classes. The pions show a monotone growth with the energy, with small discrepancies between particle and their anti-particle, mainly at lower energies, for each centrality class.
The bottom panel of Fig.~\ref{Fig.Nch.Cen.pi.pro} shows the proton case, with behavior contrary to that of pions: the proton multiplicity starts at higher values at lower \start energy and decreases as energy increases, reaching values similar to those of anti-protons at the same centrality at higher energy (200 \gev). However, the anti-proton starts with low multiplicity and increases with energy, reaching a value close to the proton multiplicity at higher energies, while a small asymmetry between protons and anti-protons increases as the energy decreases. These multiplicities grow with the centrality of the collisions. It is worth mentioning that \ags has data at lower energy where the multiplicity increases with energy. However, it is important to note that lower energy implies lower multiplicity and, consequently, a higher statistical error in the prediction.
Following the trend of multiplicity for the whole energy range analyzed (2.7-200 GeV), we observe that the multiplicity for protons increases, reaching a maximum around 7-8 \gev, and then slowly decreases with the energy, the same trend observed for $K^+, \;\; \Lambda^0, \;\; \Sigma^+$.

\begin{figure}[h]
    \centering
        \includegraphics[scale =0.45]{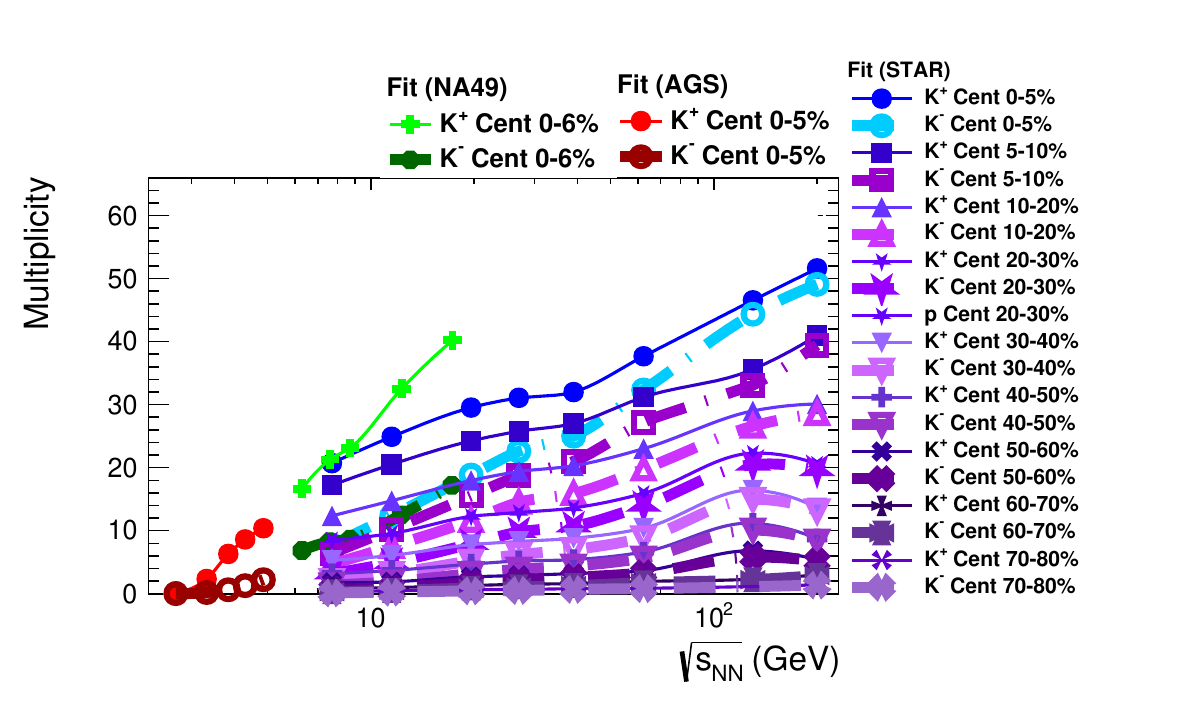}
        \includegraphics[scale =0.45]{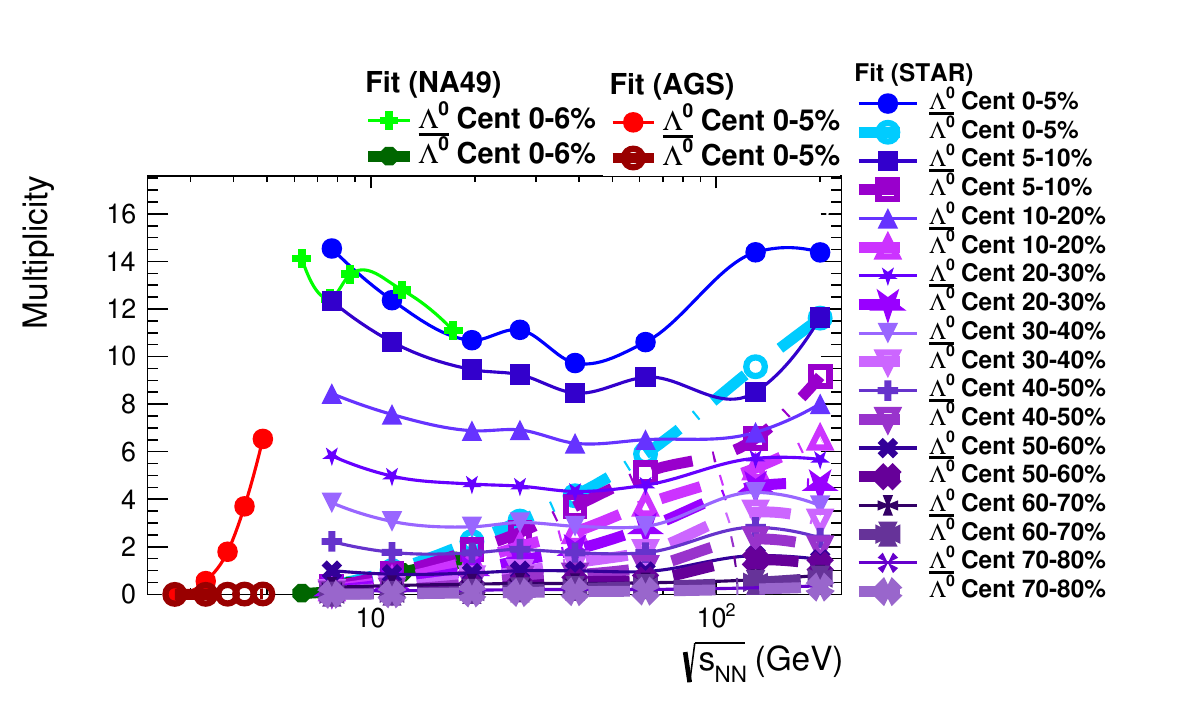}
    \caption{Multiplicity for $K^{\pm}$ (top) and $\Lambda^0, \bar{\Lambda^0}$ (bottom) for central to peripheral collisions from \start, while for \naexp and \ags only central collision are shown.}\label{Fig.Nch.Cen.LamdK}
\end{figure}

Figure~\ref{Fig.Nch.Cen.LamdK} shows the energy dependence for $K^{\pm}$ (top) and for $\Lambda$ and $\overline{\Lambda^0}$ (bottom ) for multiplicity computed for the centrality 0-6\% from \naexp and  \ags, and nine centrality classes from \start data. The behavior for $K^{\pm}$ is similar to $\pi^{\pm}$, while the $\Lambda$'s behavior looks like that of the protons, which means that the multiplicity of mesons and anti-mesons increases with energy. The baryons appear to increase, reaching a maximum and then decrease, while the anti-baryon multiplicity monotonically increases, like mesons.

\begin{figure}[h]
    \centering
    \includegraphics[scale =0.5]{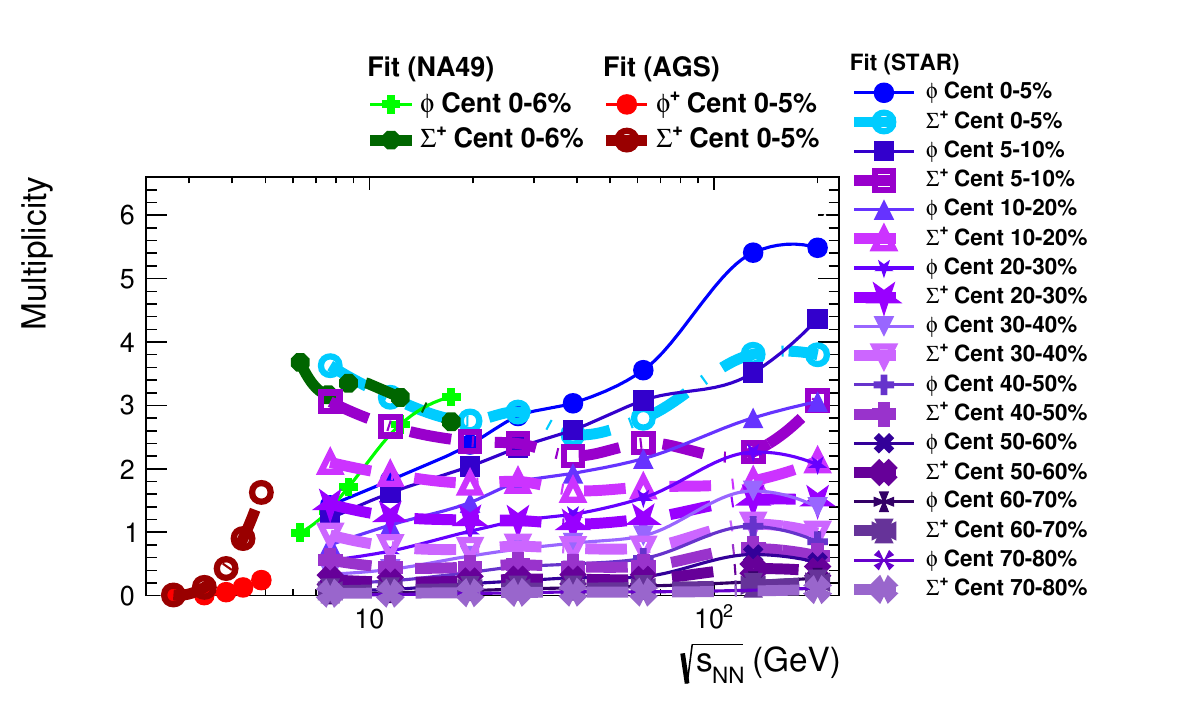}
    \includegraphics[scale =0.5]{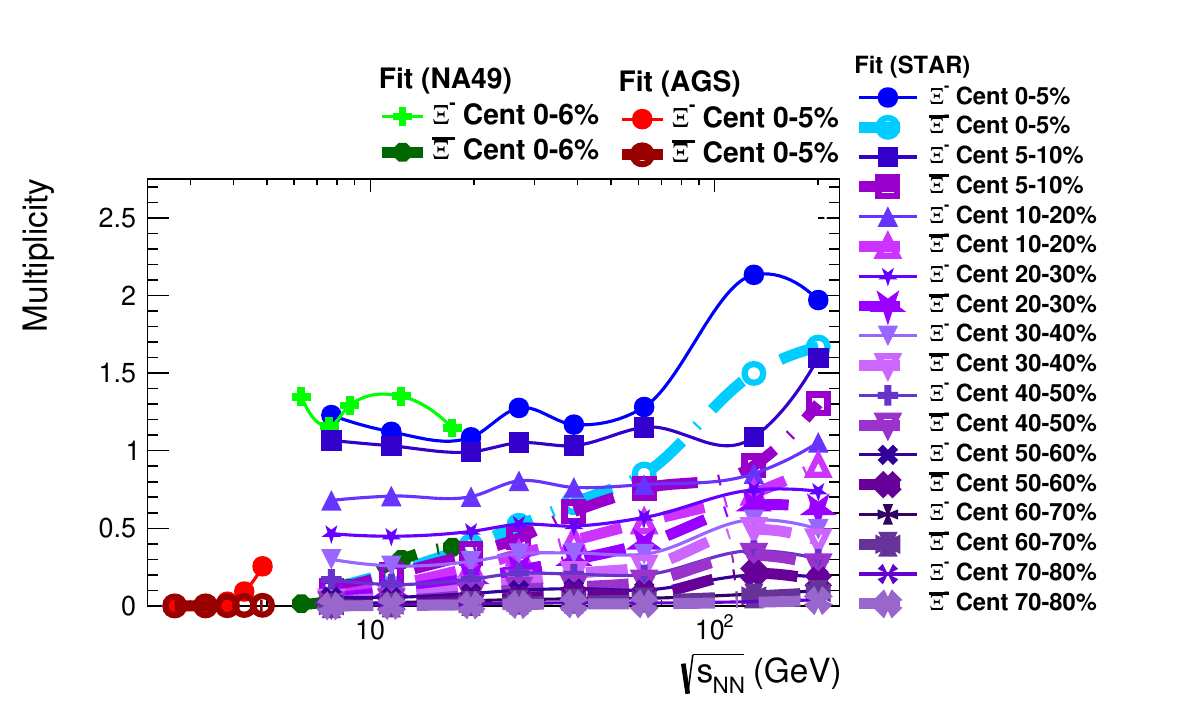}
    \includegraphics[scale =0.5]{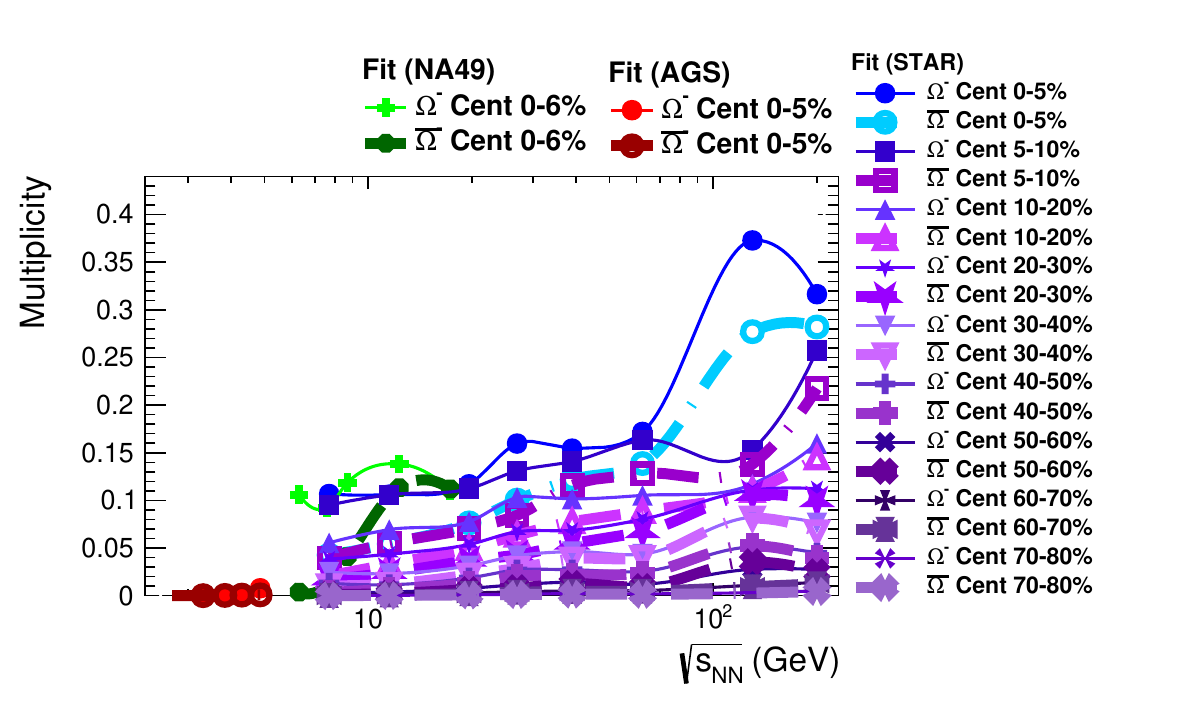}
    \caption{Multiplicity distributions as a function of the collision energy for strange and multi-strange hadrons.}\label{Fig.Nch.MultiStrange} 
\end{figure}

Once we have validated the capability of predicting multiplicity for light measured hadrons, we use the model to predict the multiplicity of multi-strange hadrons analyzed, with the results shown in Fig.~\ref{Fig.Nch.MultiStrange} for $\Omega$ (top), $\Xi$ (middle),  and $\Sigma$'s and $\phi$ (bottom). $\Xi^-$ and $\Sigma^+$ grow slowly with the energy, while the $\Xi^+$ and $\Omega$ have a slower growth. The multiplicity for the most peripheral collisions grows very slowly. Part of these facts is that it requires more energy to produce heavier hadrons.

\begin{figure}[h]
    \centering
    \includegraphics[scale=0.45]{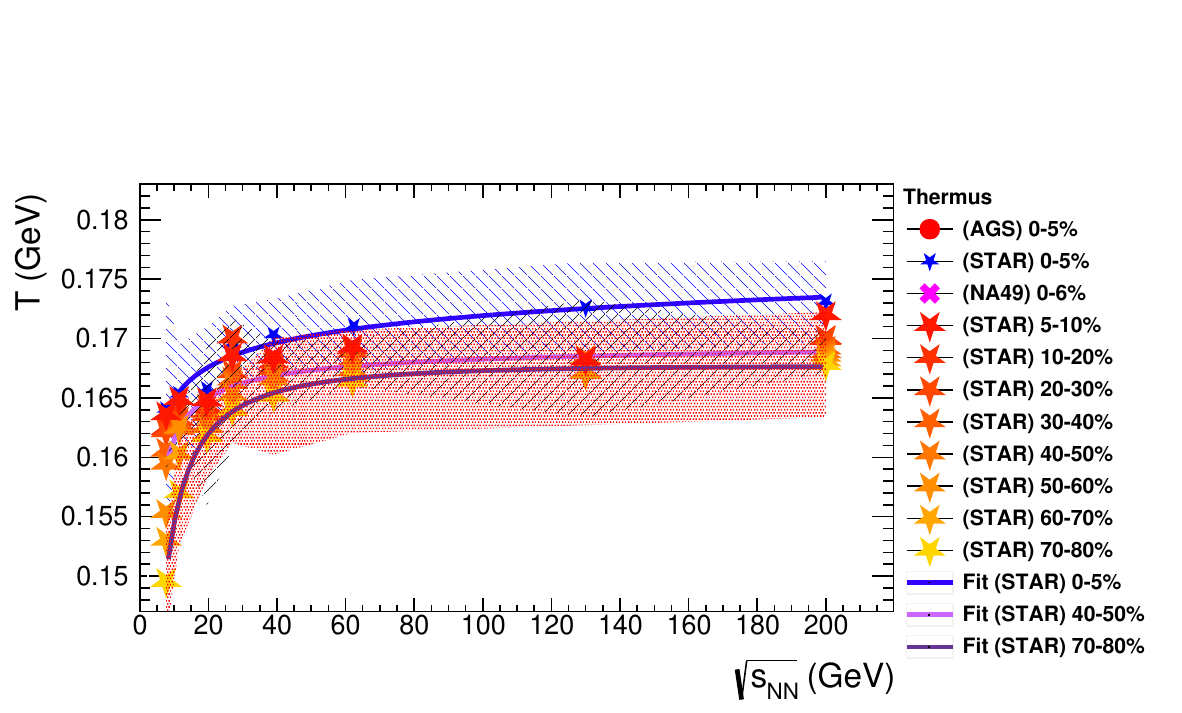}
    \caption{ Temperature predicted from \start data for nine centrality bins. Three curves correspond to a fit from (0-5)\%, (40-50)\%,  and (70-80)\% bins, from which we predict the window of the temperature from the peripheral to most central collisions.}\label{Fig.TvsCentrality}
\end{figure}

Another result predicted by this analysis is the freeze-out temperature for centrality classes from \start data, with behavior similar for all centralities but a well-defined window between the temperature from central and peripheral collisions, as shown in Fig.~\ref{Fig.TvsCentrality}.  Three centralities: (0-5)\%, (40-50)\%, and (70-80)\% are fitted by Eq.~\ref{Eq.NMu} with parameters described in table~\ref{Tab.Potentials} and Table~\ref{Chi2T}.
It should be noted that the errors arise from low multiplicity measurements, mainly at lower collision energies and peripheral collisions, and therefore cannot be reduced within the model used. The same analysis was performed on the data from \ags and \naexp, producing results similar to those of \start, but with larger error bars; therefore, these results were not plotted in the figures shown for clarity.

\begin{figure}[h]
    \centering
    \includegraphics[scale =0.45]{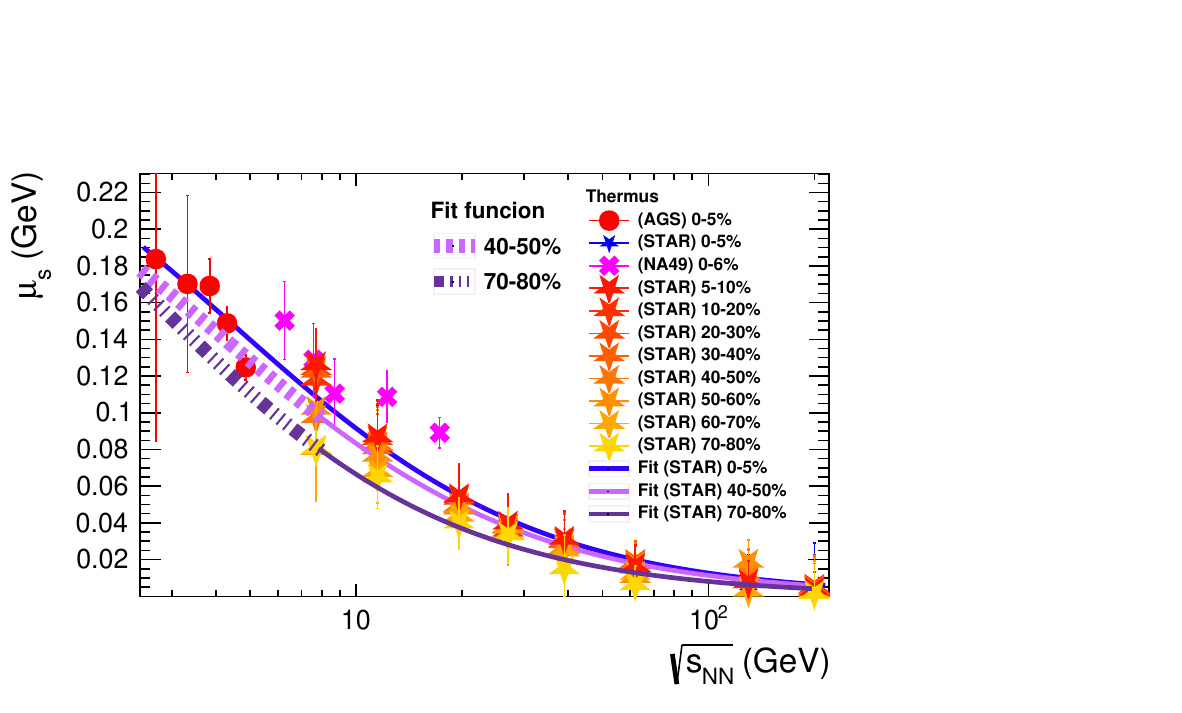} 
    \includegraphics[scale =0.45]{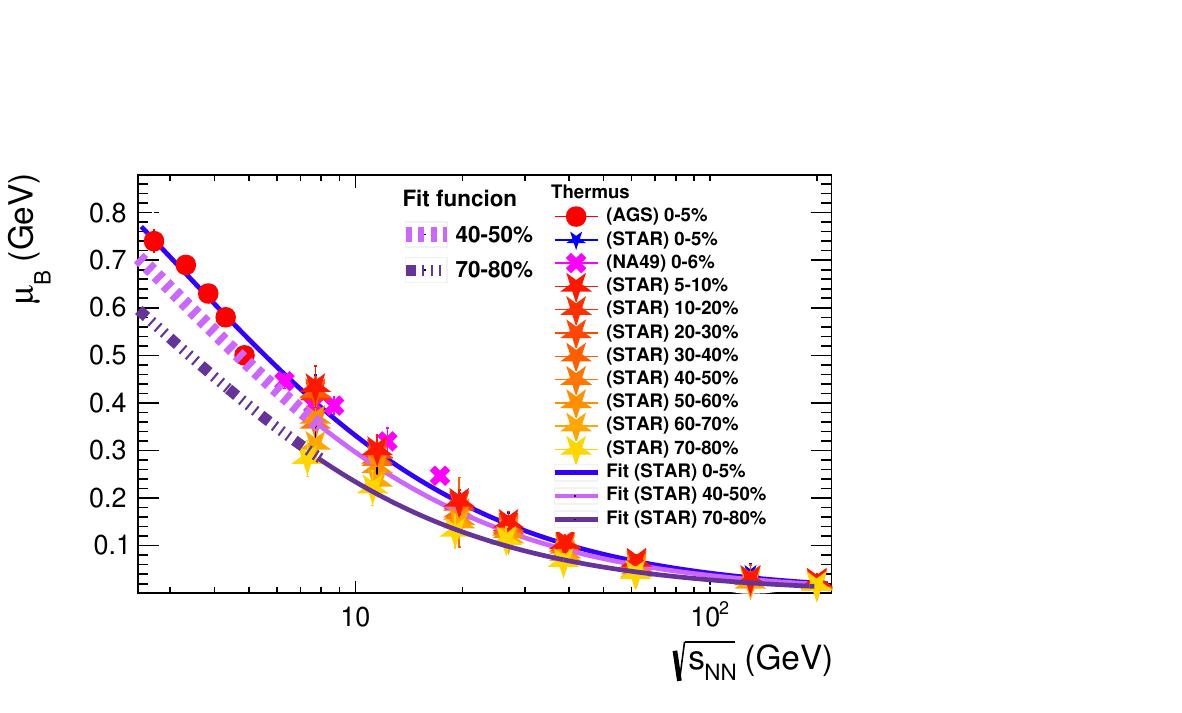} 
    \caption{Baryon chemical and strangeness potentials for nine centrality bins from \start data. Three centrality values have been fitted and then extrapolated to lower energies where there are no published results.}\label{Fig.PotentialvsCent}
\end{figure}

We also predict the energy dependence for the baryon chemical and strangeness potentials from different centrality classes, as is shown in Fig.~\ref{Fig.PotentialvsCent}, which has been fitted with Eq.~\ref{Eq.Mu} for three centralities: $(0-5)\%$, $(40-50)\%$ and $(70-80)\%$,  with parameters given in Table~\ref{Tab.Potentials} and Table~\ref{Chi2}. Since we have no data at lower energies, the fitted function was extrapolated to zero energy, allowing us to observe discrepancies, for instance, at 4 \gev $\mu_B$ is around 18\% larger in central respect to peripheral collisions, while the $\mu_s$ at the same energy shows around 3\% of differences. The general trend of this potential is an increase in discrepancies as the collision energy decreases, highlighting the importance of this analysis for future experiments at lower energies, such as \mpdnica, which should produce high enough multiplicity with error bars that allow for high precision on thermodynamic variables.

\begin{table}[htb!]
\caption{$\chi^2/ndf$ from  temperature fit for \start data, Fig.~\ref{Fig.TvsCentrality}}
    \label{Chi2T}  
    \begin{tabular}{ll}
      \hline\noalign{\smallskip}
        $T_{centrality}$     & $\chi^2/ndf$ \\  
        \noalign{\smallskip}\hline\noalign{\smallskip}
          $T (0-5\%)$& 0.8422\\
          $T(40-50\%)$ &  0.9384\\
          $T(70-80\%)$ & 1.0133 \\
  \noalign{\smallskip}\hline   
    \end{tabular}    
\end{table}

%
  \begin{table}
      \caption{ Parameters predicted for baryon chemical and strangeness potential and temperature of the  Fig.~\ref{Fig.TvsCentrality} by fitting  Eq.~\ref{Eq.NMu} (a,b,c,f) and Eq.~\ref{Eq.Mu} (d,e).}
      \label{Tab.Potentials}
      \centering{\small
  \begin{tabular}{lllll} 
    \hline\noalign{\smallskip}
    (GeV) & $a_{T}/d$     & $b_{T}/e$ & $c_{T}$ & $f_{T}$\\ 
    \noalign{\smallskip}\hline\noalign{\smallskip}
    $T_{0-5\%}$ & 0.173 & 0.143  & -0.107  &-0.157 \\
              & $\pm$ 0.002 & $\pm$ 0.04 & $\pm$ 0.032& $\pm$0.001 \\
        $\mu_{B, 0-5\%}$ & 1.373 $\pm$  0.086  & 0.315 $\pm$  0.034 & \\
        $\mu_{s, 0-5\%}$ & 0.295 $\pm$  0.024 & 0.223  $\pm$ 0.036  & \\ 
    $T_{40-50\%}$ & 0.168 & 0.205 & -1.824  &-29.4803 \\
                  &$\pm$ 0.0026 & $\pm$ 0.0630 & $\pm$ 0.3393 & $\pm$ 0.0462\\
        $\mu_{B, 40-50\%}$ & 1.950 $\pm$ 0.120 & 0.520 $\pm$ 0.038 & \\
        $\mu_{s, 40-50\%}$ & 0.280 $\pm$ 0.025 & 0.230 $\pm$ 0.095  & \\ 
    $T_{70-80\%}$ & 0.1672 & 0.3514 & -1.4411  & -6.059 \\
      & $\pm$ 0.0020 & $\pm$ 0.0297 &$\pm$ 0.9823 & $\pm$ 0.8424 \\
        $\mu_{B, 70-80\%}$ & 5.200 $\pm$1.300 & 2.65 $\pm$ 0890 & \\
        $\mu_{s, 70-80\%}$ & 5.330 $\pm$ 0.960 &  4.510 $\pm$ 1.360 & \\ 
       \noalign{\smallskip}\hline 
  \end{tabular}
  } 
\end{table}
  
\begin{table}[htb!]
\caption{$\chi^2/ndf$ from $\mu_B$ or $\mu_s$ fit for \start data, Fig.~\ref{Fig.PotentialvsCent} }
    \label{Chi2}  
    \begin{tabular}{ll}
      \hline\noalign{\smallskip}
        $\mu_B$ or $\mu_S$    & $\chi^2/NDF$ \\  
        \noalign{\smallskip}\hline\noalign{\smallskip}
             $\mu_B(0-5\%)$   &0.4353 \\
             $\mu_B(40-50\%)$ & 0.6464 \\
             $\mu_B(70-80\%)$ & 0.8224 \\
             $\mu_S(0-5\%)$ & 0.7924 \\
             $\mu_S(40-50\%)$ &  0.8034\\
             $\mu_S(70-80\%)$ & 0.8935 \\
  \noalign{\smallskip}\hline   
    \end{tabular}    
\end{table}

\begin{figure}[h!]
    \centering
    \includegraphics[scale =0.45]{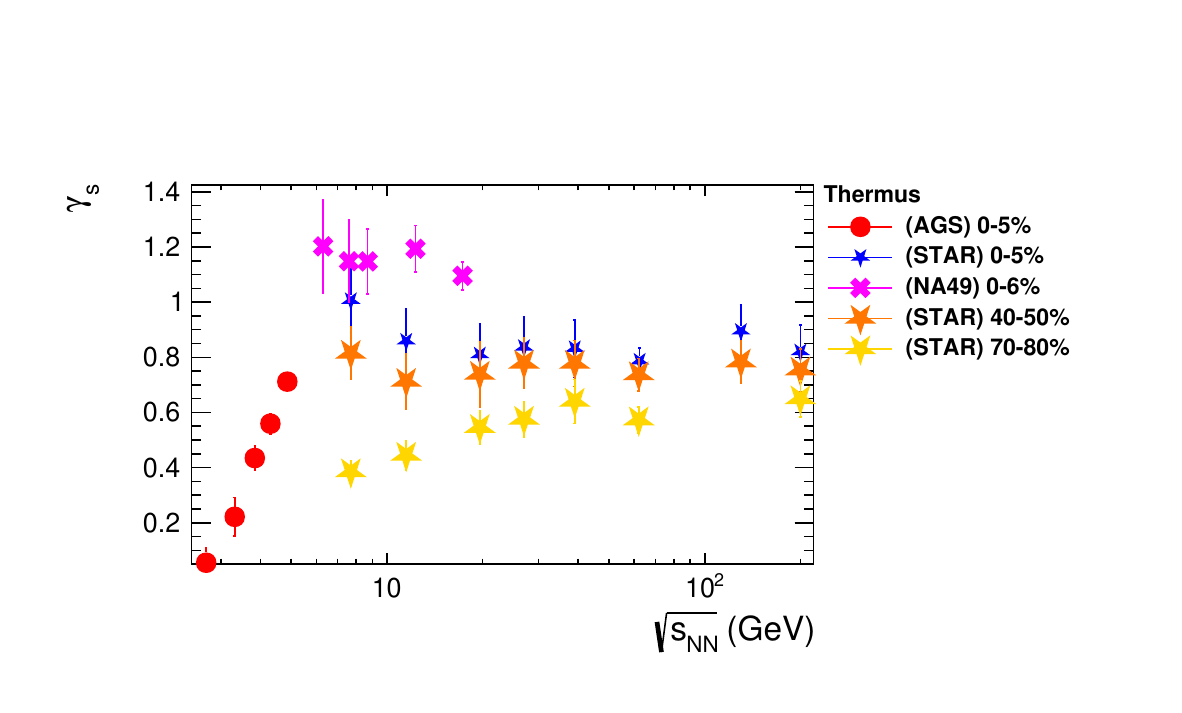}
    \includegraphics[scale =0.45]{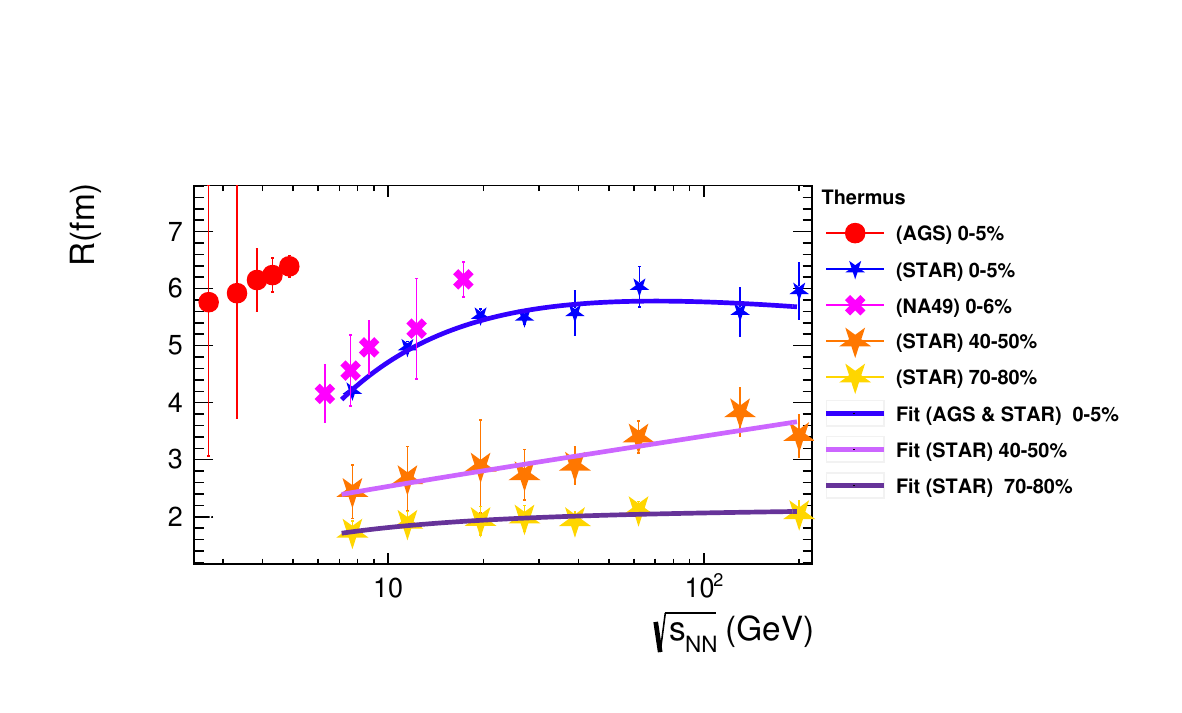}
    \caption{Energy dependence of $\gamma_{s}$ (top) and fireball radii $R$ (bottom) for four centrality classes}\label{Fig.gammacentrality}
\end{figure}

The factor $\gamma_s$ was proposed to explain the horn in the $\LambdaToPi$ ratios observed in \pbpb collision in \naexp, but not in \pp collision. Since the  \start experiment has measured the multiplicity in centrality classes, we use it to obtain the $\gamma_s$ distributions shown in Fig.~\Ref{Fig.gammacentrality}. The $\gamma_s$ distribution for peripheral collisions starts from lower values at lower energy and increases with energy. However, when going from peripheral to central collisions, a maximum appears in the distribution around 7-8 \gev. The behavior in $\gamma_s$ for central collisions is associated with an increase in the multiplicity of baryons, as previously presented, as well as with the physical processes that occur during the evolution of systems. Furthermore, the centrality classes are also reflected in the size of systems where the radii can be predicted; the bottom panel of Fig.~\Ref{Fig.gammacentrality} shows the radii for different centrality classes, where a drastic change in the energy 7-8 \gev is observed. It is important to note that at lower energies (\ags) the radii are larger, but with considerably larger uncertainty. These differences could come from different selection processes used for hadrons measured in \ags compared to \start and \naexp, discussed in ~\ref{data}, as well as the low number of identified hadrons and the low multiplicity for each species. Peripheral collisions produce a fireball radius of around 2 $fm$, which is consistent with proton-proton collisions.

\begin{figure}[h!]
  \centering
  \includegraphics[scale=0.55]{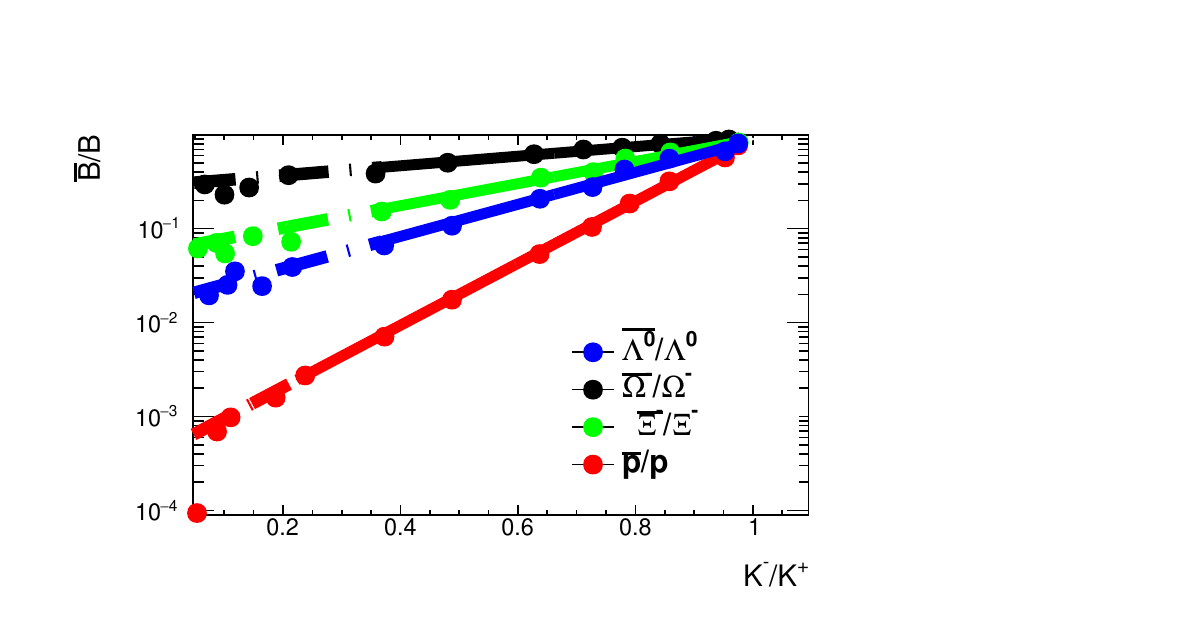}
    \caption{$\bar{p}/p$,$\bar{\Lambda^0}/\Lambda^0$, $\bar{\Xi^{-}}/\Xi^+$ y $\bar{\Omega^{-}}/\Omega^+$ vs $K^{-}/K^+$}\label{Fig.RatiosStrangeness}
\end{figure}

The strange hadron production investigated in the deconfinement phase has also been studied by Das and Hwa using the recombination model and applied to heavy-ion collisions by Wang~\cite{Wang:2013duu}. Those studies compute the baryon anti-baryon ($\bar{B}/B$) and suggest a universal behavior as a function of $K^-/K^+$. To explore this idea with thermal model and experimental multiplicity as input, we compute the ratios  $\bar{p}/p$, $\bar{\Lambda^0}/\Lambda^0$, $\bar{\Xi^{-}}/\Xi^+$ and $\bar{\Omega^{-}}/\Omega^+$, all of them plotted as a function of the $K^{-}/K^+$ ratio, are shown in Fig.~\ref{Fig.RatiosStrangeness}. The $\bar{B}/B$ ratio can be parametrized by  Eq.~\ref{Eq.Ratios} with the parameters given in Table.~\ref{Table.Ratios}.

\begin{equation}\label{Eq.Ratios}
    \bar{B}/B = a e^{b+c(\sqrt{s_{NN}})}
\end{equation}

\begin{table}[h!]
 \caption{Parameters from a fit of baryon antibaryon ratios with Eq.~\ref{Eq.Ratios}}
    \label{Table.Ratios}
    \begin{tabular}{llll} 
      \hline\noalign{\smallskip}
      & a & b & c\\  
     \noalign{\smallskip}\hline\noalign{\smallskip} 
        $\Omega^-/\bar{\Omega^+}$ & 0.663 & -0.821 & 1.174\\  
        $\Xi^-/\bar{\Xi^+}$ & 0.361 &  -1.805& 2.708\\
        $\Lambda^0/\bar{\Lambda^0}$ & 0.181 & -2.362 &  3.933\\
     $p/\bar{p}$& 0.012 & -3.300 & 7.596\\ 
     \noalign{\smallskip}\hline
    \end{tabular}   
\end{table}

The figure displays a continuous line corresponding to the fit and dashed lines representing its extrapolations. The extrapolated values correspond to a region with low statistics and large fluctuations, but in general, Eq.~\ref{Eq.Ratios} describe the ratios reasonably well. It is important to note the convergence of the ratios when the $K^-/K^+$ ratio approaches one.

\section{Skewness in QVdW-HRG model}

Once we have the partition function from the statistical model, we can compute the system's properties, including the susceptibility and its ratios, such as those associated with strangeness ($\chi_s$) between third and second order $\chi^{(3)}_s$/$\chi^{(2)}_s$, known as skewness ($S\sigma$). By fitting the hadron multiplicities in Table 1, one can determine the thermal parameters and then compute the partition function from which we obtain the skewness, which is discussed in the rest of this section.\\

Applying the \qvdw equation of state to describe nuclear matter, one can predict the existence of a first-order liquid-gas phase transition and a critical point~\cite{Vovchenko:2015pya}. In our case, we analyze the experimental multiplicities to obtain and compare the freeze-out lines in the ($\mu_B, T$) plane, using the equations of state: \gce with ideal \hrg (\gce-\hrg). As shown in Fig.~\ref{Fig.muvsT}, the temperature-dependent parametric function shows differences that increase as energy decreases. 

\begin{figure}[h!]
    \includegraphics[scale=0.45]{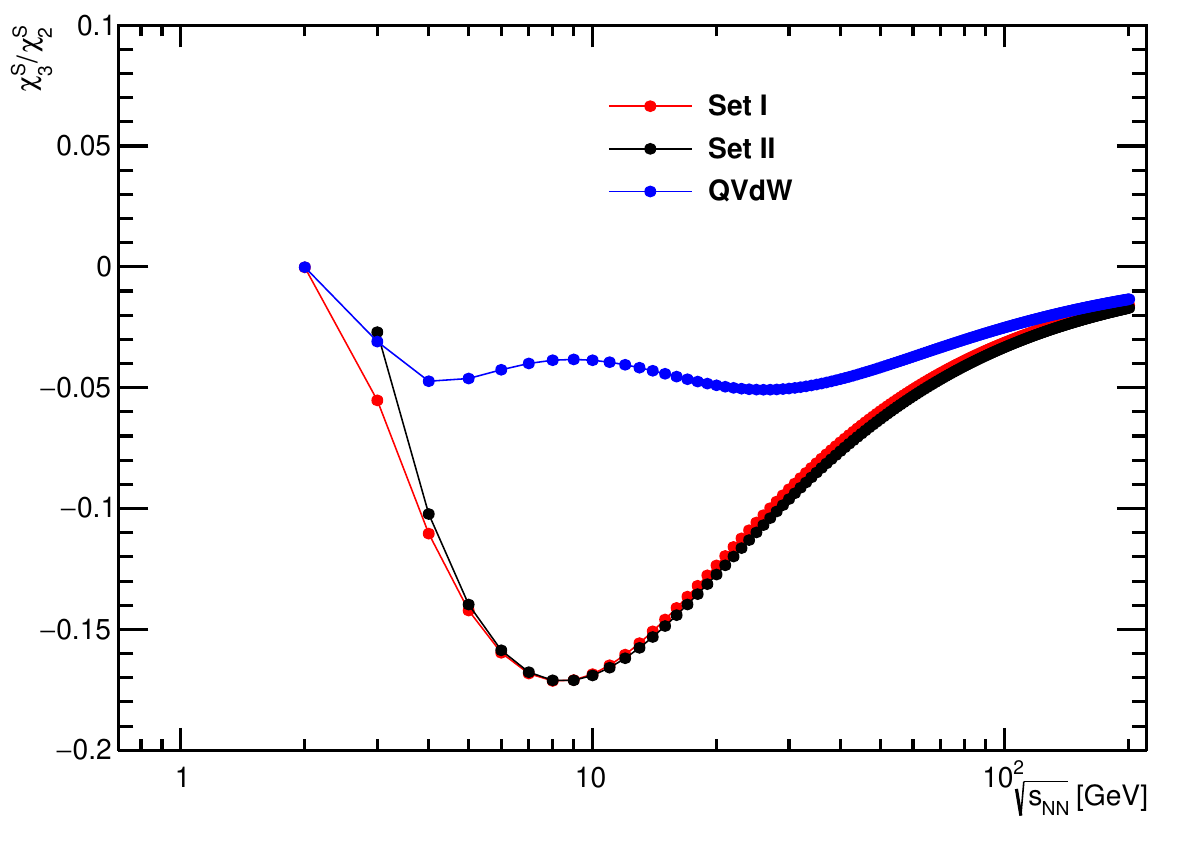}
    \caption{ Skewness for the strangeness using \gce-\hrg  with two sets of parameters as described in the text,  and \gce-\qvdw model.}\label{Fig.freeze-out}
\end{figure}

The skewness is shown in Fig.\ref{Fig.freeze-out} for three parametric functions of the freeze-out temperature: two parametric Ideal \hrg (with legends Set I and Set II) and one for \qvdw. The results show a significant discrepancy between the two models, even for \hrg obtained along the two freeze-out lines, as shown in Fig.~\ref{Fig.muvsT}, where differences in energy below 10 GeV are observed.
Low collision energy has interest because there are the most significant differences between skewness and kurtosis of net-electric, net-baryonic and net-strangeness charge fluctuations~\cite{Poberezhnyuk:2019pxs} computed with \qvdw-\hrg and \hrg models. However, the hypothesis of thermodynamic equilibrium at freeze-out is admisible at the highest energies, and it is less probable at lower collision energy~\cite{Gupta:2022phu}.


The skewness calculated with the \qvdw-\hrg model shows a non-monotonic behavior with collision energy.  Furthermore, it depends on the parameters of the model, such as strangeness and electric over baryon charge (Q/B =0), for which it is known that the model exhibits a critical point at $\mu_B=914.5$ MeV and $T_c =19.48$ MeV, widely discussed~\cite{Poberezhnyuk:2019pxs}. 
Although skewness is model dependent, it has been related to experimental event-by-event fluctuations of conserved charged~\cite{Karsch:2010ck}, but these analyses are outside the scope of this work.


\section{Conclusions}\label{conclusion}
Multiplicity distributions for identified hadrons produced in \pbpb collisions in the \naexp experiment,  and  \auau in \ags and \start experiments in an energy range between 2.7-200 \gev were analyzed using the \thermal model. We predict the  thermodynamic properties of the system considered as a fireball created in heavy ion collisions.

In the first step, experimental multiplicity was used as input, yielding the same multiplicity as the output, allowing the model to be validated. The obtained hadron multiplicity corresponds to those used as input, as well as others, such as multi-strange hadrons. The complete list of analyzed hadrons is $\pi's$, $K's$, $p$, $\Lambda's$, $\phi's$, $\Sigma's$, $\Xi's$, $\Omega's$, and their anti-particles. \\
The model enables the construction of the partition function, from which it is possible to predict thermodynamic properties of the system created in heavy-ion collisions, which are summarized as follows:\\

\begin{itemize}
\item The obtained hadron multiplicity was used to compute ratios, such as \KaonToPionPlus, \KaonToPionMinos, and  \LambdaToPi, which show good agreement with the experimental results. Predictions for the \OmegaToOmegaPM and \ALambdaLambda ratios as a function of the collision energy are reported.
\item The freeze-out temperature and baryon chemical and strangeness potentials were predicted using a more precise parametric function to improve the agreement between the data and the model. In this context, the predicted temperature is 8\% higher than that reported by previous measurements, which can be explained by the free/fixed parameters in the fit. Because temperature is a function of the baryon chemical potential, any change in one of them produces a modification in the other.
\item Volume and the strangeness suppression factor were also analyzed, finding agreement between the results of \start and \naexp. However, the \ags results show considerable uncertainty due to the very low multiplicity reported.
\item The anti-Baryon to Baryon ratios as a function of $K^{-}/K^{+}$ were calculated, finding a universal parametrization as a function of collision energy. These results agree with the parton recombination model.
\item A more detailed analysis was performed using the multiplicity by centrality classes reported by the \start experiment. We predict
multiplicity for multi-strange hadrons. We found the freeze-out temperature in the range of 167-177 \mev, limits imposed by peripheral and most central collisions, respectively.
\item The baryon chemical and strangeness potentials computed for centrality classes show a decrease in peripheral collisions with respect to central ones when the energy decreases.
\item  The skewness associated with the strangeness was calculated along the freeze-out line obtained from a fit of experimental yields at different energies for the HRG and QVdW-HRG models, revealing clear differences between the two, as expected. It is more important to point out the differences in energy below 10 GeV, between the skewness of the HRG computed along the two freeze-out lines.

\end{itemize}


The analysis presented here shows that the thermodynamic properties of the system formed in heavy-ion collisions change significantly as collision energy decreases, making them of interest at FAIR and NICA energies.

\section*{Acknowledgments}
This work was funded by  DGAPA-PAPIIT IG100826  and SECIHTI CBF-2026-465  projects. The author thanks R. Garcia-Formenti for the discussion and his comments.


\bibliographystyle{unsrt}  
\bibliography{Ref}

@article{Aoki:2006we,
    author = "Aoki, Y. and Endrodi, G. and Fodor, Z. and Katz, S. D. and Szabo, K. K.",
    title = "{The Order of the quantum chromodynamics transition predicted by the standard model of particle physics}",
    eprint = "hep-lat/0611014",
    archivePrefix = "arXiv",
    doi = "10.1038/nature05120",
    journal = "Nature",
    volume = "443",
    pages = "675--678",
    year = "2006"
}

@article{Cheng:2007jq,
    author = "Cheng, M. and others",
    title = "{The QCD equation of state with almost physical quark masses}",
    eprint = "0710.0354",
    archivePrefix = "arXiv",
    primaryClass = "hep-lat",
    reportNumber = "BNL-NT-07-38, BI-TP-2007-20, CU-TP-1181",
    doi = "10.1103/PhysRevD.77.014511",
    journal = "Phys. Rev. D",
    volume = "77",
    pages = "014511",
    year = "2008"
}

@article{Akiba:2015jwa,
    author = "Akiba, Yasuyuki and others",
    title = "{The Hot QCD White Paper: Exploring the Phases of QCD at RHIC and the LHC}",
    eprint = "1502.02730",
    archivePrefix = "arXiv",
    primaryClass = "nucl-ex",
    month = "2",
    year = "2015"
}

@article{Du:2024wjm,
    author = "Du, Lipei and Sorensen, Agnieszka and Stephanov, Mikhail",
    title = "{The QCD phase diagram and Beam Energy Scan physics: A theory overview}",
    eprint = "2402.10183",
    archivePrefix = "arXiv",
    primaryClass = "nucl-th",
    reportNumber = "INT-PUB-24-017",
    doi = "10.1142/9789811294679_0007",
    journal = "Int. J. Mod. Phys. E",
    volume = "33",
    number = "07",
    pages = "2430008",
    year = "2024"
}

@article{Rafelski:1982pu,
    author = "Rafelski, Johann and Muller, Berndt",
    title = "{Strangeness Production in the Quark - Gluon Plasma}",
    reportNumber = "Print-82-0048 (FRANKFURT)",
    doi = "10.1103/PhysRevLett.48.1066",
    journal = "Phys. Rev. Lett.",
    volume = "48",
    pages = "1066",
    year = "1982",
    note = "[Erratum: Phys.Rev.Lett. 56, 2334 (1986)]"
}

@article{NA49:2004iqm,
    author = "Gazdzicki, M. and others",
    editor = "Ritter, Hans Georg and Wang, Xin-Nian",
    collaboration = "NA49",
    title = "{Report from NA49}",
    eprint = "nucl-ex/0403023",
    archivePrefix = "arXiv",
    doi = "10.1088/0954-3899/30/8/008",
    journal = "J. Phys. G",
    volume = "30",
    pages = "S701--S708",
    year = "2004"
}

@article{Cleymans:2004hj,
    author = "Cleymans, J. and Oeschler, H. and Redlich, K. and Wheaton, S.",
    title = "{Transition from baryonic to mesonic freeze-out}",
    eprint = "hep-ph/0411187",
    archivePrefix = "arXiv",
    doi = "10.1016/j.physletb.2005.03.074",
    journal = "Phys. Lett. B",
    volume = "615",
    pages = "50--54",
    year = "2005"
}

@inproceedings{Becattini:2009sc,
    author = "Becattini, F.",
    title = "{An Introduction to the Statistical Hadronization Model}",
    booktitle = "{International School on Quark-Gluon Plasma and Heavy Ion Collisions: past, present, future}",
    eprint = "0901.3643",
    archivePrefix = "arXiv",
    primaryClass = "hep-ph",
    month = "1",
    year = "2009"
}

@article{Oeschler:2006wkw,
    author = "Oeschler, H. and Cleymans, J. and Redlich, K. and Wheaton, S.",
    title = "{Transition from baryon to meson-dominated freeze out: early decoupling around 30-A-GeV?}",
    eprint = "nucl-th/0701080",
    archivePrefix = "arXiv",
    doi = "10.1088/0954-3899/32/12/S28",
    journal = "J. Phys. G",
    volume = "32",
    pages = "S223--S229",
    year = "2006"
}

@article{Ayala:2023lnl,
    author = "Ayala, Alejandro and Bietenholz, Wolfgang and Cuautle, Eleazar and Forment\'\i{}, Rodrigo Garc\'\i{}a and Guzm\'an, Rodrigo",
    title = "{Searching for the Baryon-to-Meson Transition Region with the MPD at NICA}",
    eprint = "2305.02455",
    archivePrefix = "arXiv",
    primaryClass = "hep-ph",
    doi = "10.1134/S1063778823050071",
    journal = "Phys. Atom. Nucl.",
    volume = "86",
    number = "5",
    pages = "901--907",
    year = "2023"
}

@article{Cleymans:2004pp,
    author = "Cleymans, J. and Kampfer, Burkhard and Kaneta, M. and Wheaton, S. and Xu, N.",
    title = "{Centrality dependence of thermal parameters deduced from hadron multiplicities in Au + Au collisions at s(NN)**(1/2) = 130-GeV}",
    eprint = "hep-ph/0409071",
    archivePrefix = "arXiv",
    doi = "10.1103/PhysRevC.71.054901",
    journal = "Phys. Rev. C",
    volume = "71",
    pages = "054901",
    year = "2005"
}

@article{Vovchenko:2015vxa,
    author = "Vovchenko, V. and Anchishkin, D. V. and Gorenstein, M. I.",
    title = "{Van der Waals Equation of State with Fermi Statistics for Nuclear Matter}",
    eprint = "1504.01363",
    archivePrefix = "arXiv",
    primaryClass = "nucl-th",
    doi = "10.1103/PhysRevC.91.064314",
    journal = "Phys. Rev. C",
    volume = "91",
    number = "6",
    pages = "064314",
    year = "2015"
}

@article{Monnai:2019hkn,
    author = {Monnai, Akihiko and Schenke, Bj\"orn and Shen, Chun},
    title = "{Equation of state at finite densities for QCD matter in nuclear collisions}",
    eprint = "1902.05095",
    archivePrefix = "arXiv",
    primaryClass = "nucl-th",
    reportNumber = "KEK-TH-2106",
    doi = "10.1103/PhysRevC.100.024907",
    journal = "Phys. Rev. C",
    volume = "100",
    number = "2",
    pages = "024907",
    year = "2019"
}

@article{Poberezhnyuk:2018mwt,
    author = "Poberezhnyuk, Roman and Vovchenko, Volodymyr and Gorenstein, Mark I. and Stoecker, Horst",
    title = "{Noncongruent phase transitions in strongly interacting matter within the quantum van der Waals model}",
    eprint = "1810.07640",
    archivePrefix = "arXiv",
    primaryClass = "hep-ph",
    doi = "10.1103/PhysRevC.99.024907",
    journal = "Phys. Rev. C",
    volume = "99",
    number = "2",
    pages = "024907",
    year = "2019"
}

@article{BRAHMS:2004adc,
    author = "{BRAHMS Collaboration}",
    title = "{Quark–gluon plasma and color glass condensate at RHIC? The perspective from the BRAHMS experiment}",
issn = {0375-9474},
doi = {https://doi.org/10.1016/j.nuclphysa.2005.02.130},
    journal = "Nucl. Phys. A",
    volume = "757",
number = {1},
pages = {1-27},
    year = "2005"
}

@article{PHOBOS:2004zne,
    author = "{PHOBOS Collaboration}",
    title = "{The PHOBOS perspective on discoveries at RHIC}",
issn = {0375-9474},
doi = {https://doi.org/10.1016/j.nuclphysa.2005.03.084},
    journal = "Nucl. Phys. A",
    volume = "757",
number = {1},
pages = {28-101},
    year = "2005"
}

@article{STAR:2005gfr,
    author = "{STAR Collaboration}",
    title = "{Experimental and theoretical challenges in the search for the quark–gluon plasma: The STAR Collaboration's critical assessment of the evidence from RHIC collisions}",
issn = {0375-9474},
doi = {https://doi.org/10.1016/j.nuclphysa.2005.03.085},
    journal = "Nucl. Phys. A",
    volume = "757",
number = {1},
pages = {102-183},
    year = "2005"
}

@article{PHENIX:2004vcz,
    author = "{PHENIX Collaboration}",
    title = "{Formation of dense partonic matter in relativistic nucleus–nucleus collisions at RHIC: Experimental evaluation by the PHENIX Collaboration}",
issn = {0375-9474},
doi = {https://doi.org/10.1016/j.nuclphysa.2005.03.086},
    journal = "Nucl. Phys. A",
    volume = "757",
number = {1},
pages = {184-283},
    year = "2005"
}

@article{STAR:2003ryp,
    author = "Adams, J. and others",
    collaboration = "STAR",
    title = "{Rapidity and centrality dependence of proton and anti-proton production from Au-197 + Au-197 collisions at (S(NN))**(1/2)) = 130-GeV}",
    eprint = "nucl-ex/0306029",
    archivePrefix = "arXiv",
    doi = "10.1103/PhysRevC.70.041901",
    journal = "Phys. Rev. C",
    volume = "70",
    pages = "041901",
    year = "2004"
}

@article{STAR:2003jwm,
    author = "Adams, J. and others",
    collaboration = "STAR",
    title = "{Identified particle distributions in pp and Au+Au collisions at s(NN)**(1/2) = 200 GeV}",
    eprint = "nucl-ex/0310004",
    archivePrefix = "arXiv",
    doi = "10.1103/PhysRevLett.92.112301",
    journal = "Phys. Rev. Lett.",
    volume = "92",
    pages = "112301",
    year = "2004"
}

@article{STAR:2010dor,
    author = "Aggarwal, M. M. and others",
    collaboration = "STAR",
    title = "{Scaling properties at freeze-out in relativistic heavy ion collisions}",
    eprint = "1008.3133",
    archivePrefix = "arXiv",
    primaryClass = "nucl-ex",
    doi = "10.1103/PhysRevC.83.034910",
    journal = "Phys. Rev. C",
    volume = "83",
    pages = "034910",
    year = "2011"
}

@article{STAR:2017sal,
    author = "Adamczyk, L. and others",
    collaboration = "STAR",
    title = "{Bulk Properties of the Medium Produced in Relativistic Heavy-Ion Collisions from the Beam Energy Scan Program}",
    eprint = "1701.07065",
    archivePrefix = "arXiv",
    primaryClass = "nucl-ex",
    doi = "10.1103/PhysRevC.96.044904",
    journal = "Phys. Rev. C",
    volume = "96",
    number = "4",
    pages = "044904",
    year = "2017"
}

@article{STAR:2008med,
    author = "Abelev, B. I. and others",
    collaboration = "STAR",
    title = "{Systematic Measurements of Identified Particle Spectra in $p p, d^+$ Au and Au+Au Collisions from STAR}",
    eprint = "0808.2041",
    archivePrefix = "arXiv",
    primaryClass = "nucl-ex",
    doi = "10.1103/PhysRevC.79.034909",
    journal = "Phys. Rev. C",
    volume = "79",
    pages = "034909",
    year = "2009"
}

@article{Andronic:2005yp,
    author = "Andronic, A. and Braun-Munzinger, P. and Stachel, J.",
    title = "{Hadron production in central nucleus-nucleus collisions at chemical freeze-out}",
    eprint = "nucl-th/0511071",
    archivePrefix = "arXiv",
    doi = "10.1016/j.nuclphysa.2006.03.012",
    journal = "Nucl. Phys. A",
    volume = "772",
    pages = "167--199",
    year = "2006"
}

@article{E802:1996owm,
    author = "Akiba, Y. and others",
    editor = "Braun-Munzinger, P. and Specht, H. J. and Stock, R. and Stoecker, Horst",
    collaboration = "E802",
    title = "{Particle production in Au + Au collisions from BNL E866}",
    doi = "10.1016/S0375-9474(96)00350-8",
    journal = "Nucl. Phys. A",
    volume = "610",
    pages = "139C--152C",
    year = "1996"
}

@article{E-802:1998xum,
    author = "Ahle, L. and others",
    collaboration = "E-802",
    title = "{Particle production at high baryon density in central Au+Au reactions at 11.6A GeV/c}",
    doi = "10.1103/PhysRevC.57.R466",
    journal = "Phys. Rev. C",
    volume = "57",
    number = "2",
    pages = "R466--R470",
    year = "1998"
}

@article{E866:1999ktz,
    author = "Ahle, L. and others",
    collaboration = "E866, E917",
    title = "{Excitation function of K+ and pi+ production in Au + Au reactions at 2/A-GeV to 10/A-GeV}",
    eprint = "nucl-ex/9910008",
    archivePrefix = "arXiv",
    doi = "10.1016/S0370-2693(00)00037-X",
    journal = "Phys. Lett. B",
    volume = "476",
    pages = "1--8",
    year = "2000"
}

@article{E877:1999qdc,
    author = "Barrette, J. and others",
    collaboration = "E877",
    title = "{Proton and pion production in Au + Au collisions at 10.8A-GeV/c}",
    eprint = "nucl-ex/9910004",
    archivePrefix = "arXiv",
    reportNumber = "MG99-PHY02",
    doi = "10.1103/PhysRevC.62.024901",
    journal = "Phys. Rev. C",
    volume = "62",
    pages = "024901",
    year = "2000"
}

@article{E802:1999hit,
    author = "Ahle, L. and others",
    collaboration = "E802",
    title = "{Proton and deuteron production in Au + Au reactions at 11.6/A-GeV/c}",
    doi = "10.1103/PhysRevC.60.064901",
    journal = "Phys. Rev. C",
    volume = "60",
    pages = "064901",
    year = "1999"
}

@article{E866:2000dog,
    author = "Ahle, L and others",
    collaboration = "E866, E917",
    title = "{An Excitation function of K- and K+ production in Au + Au reactions at the AGS}",
    eprint = "nucl-ex/0008010",
    archivePrefix = "arXiv",
    doi = "10.1016/S0370-2693(00)00916-3",
    journal = "Phys. Lett. B",
    volume = "490",
    pages = "53--60",
    year = "2000"
}

@article{E895:2001zms,
    author = "Klay, J. L. and others",
    collaboration = "E895",
    title = "{Longitudinal flow from 2-A-GeV to 8-A-GeV Au+Au collisions at the Brookhaven AGS}",
    eprint = "nucl-ex/0111006",
    archivePrefix = "arXiv",
    doi = "10.1103/PhysRevLett.88.102301",
    journal = "Phys. Rev. Lett.",
    volume = "88",
    pages = "102301",
    year = "2002"
}

@article{NA49:2002pzu,
    author = "Afanasiev, S. V. and others",
    collaboration = "NA49",
    title = "{Energy dependence of pion and kaon production in central Pb + Pb collisions}",
    eprint = "nucl-ex/0205002",
    archivePrefix = "arXiv",
    doi = "10.1103/PhysRevC.66.054902",
    journal = "Phys. Rev. C",
    volume = "66",
    pages = "054902",
    year = "2002"
}

@article{NA49:2004mrq,
    author = "Anticic, T. and others",
    collaboration = "NA49",
    title = "{Energy and centrality dependence of deuteron and proton production in Pb + Pb collisions at relativistic energies}",
    doi = "10.1103/PhysRevC.69.024902",
    journal = "Phys. Rev. C",
    volume = "69",
    pages = "024902",
    year = "2004"
}

@article{NA49:2006gaj,
    author = "Alt, C. and others",
    collaboration = "NA49",
    title = "{Energy and centrality dependence of anti-p and p production and the anti-Lambda/anti-p ratio in Pb+Pb collisions between 20/A-GeV and 158/A-Gev}",
    doi = "10.1103/PhysRevC.73.044910",
    journal = "Phys. Rev. C",
    volume = "73",
    pages = "044910",
    year = "2006"
}

@article{NA49:2007stj,
    author = "Alt, C. and others",
    collaboration = "NA49",
    title = "{Pion and kaon production in central Pb + Pb collisions at 20-A and 30-A-GeV: Evidence for the onset of deconfinement}",
    eprint = "0710.0118",
    archivePrefix = "arXiv",
    primaryClass = "nucl-ex",
    doi = "10.1103/PhysRevC.77.024903",
    journal = "Phys. Rev. C",
    volume = "77",
    pages = "024903",
    year = "2008"
}

@article{PHENIX:2004vdg,
    author = "Adler, S. S. and others",
    collaboration = "PHENIX",
    title = "{Systematic studies of the centrality and s(NN)**(1/2) dependence of the d E(T) / d eta and d (N(ch) / d eta in heavy ion collisions at mid-rapidity}",
    eprint = "nucl-ex/0409015",
    archivePrefix = "arXiv",
    doi = "10.1103/PhysRevC.71.034908",
    journal = "Phys. Rev. C",
    volume = "71",
    pages = "034908",
    year = "2005",
    note = "[Erratum: Phys.Rev.C 71, 049901 (2005)]"
}

@article{Randrup:2006nr,
    author = "Randrup, Jorgen and Cleymans, Jean",
    title = "{Maximum freeze-out baryon density in nuclear collisions}",
    eprint = "hep-ph/0607065",
    archivePrefix = "arXiv",
    reportNumber = "LBNL-60923",
    doi = "10.1103/PhysRevC.74.047901",
    journal = "Phys. Rev. C",
    volume = "74",
    pages = "047901",
    year = "2006"
}

@article{Vovchenko:2019pjl,
    author = "Vovchenko, Volodymyr and Stoecker, Horst",
    title = "{Thermal-FIST: A package for heavy-ion collisions and hadronic equation of state}",
    eprint = "1901.05249",
    archivePrefix = "arXiv",
    primaryClass = "nucl-th",
    doi = "10.1016/j.cpc.2019.06.024",
    journal = "Comput. Phys. Commun.",
    volume = "244",
    pages = "295--310",
    year = "2019"
}

@article{Vovchenko:2015pya,
    author = "Vovchenko, V. and Anchishkin, D. V. and Gorenstein, M. I. and Poberezhnyuk, R. V.",
    title = "{Scaled variance, skewness, and kurtosis near the critical point of nuclear matter}",
    eprint = "1506.05763",
    archivePrefix = "arXiv",
    primaryClass = "nucl-th",
    doi = "10.1103/PhysRevC.92.054901",
    journal = "Phys. Rev. C",
    volume = "92",
    number = "5",
    pages = "054901",
    year = "2015"
}

@article{MPD:2025jzd,
    author = "Abdulin, R. and others",
    collaboration = "MPD",
    title = "{MPD physics performance studies in Bi+Bi collisions at {\ensuremath{\sqrt{}}}sNN = 9.2 GeV}",
    eprint = "2503.21117",
    archivePrefix = "arXiv",
    primaryClass = "nucl-ex",
    doi = "10.31349/RevMexFis.71.041201",
    journal = "Rev. Mex. Fis.",
    volume = "71",
    number = "4",
    pages = "041201",
    year = "2025"
}

@article{Floris:2014pta,
    author = "Floris, Michele",
    editor = "Braun-Munzinger, Peter and Friman, Bengt and Stachel, Johanna",
    title = "{Hadron yields and the phase diagram of strongly interacting matter}",
    eprint = "1408.6403",
    archivePrefix = "arXiv",
    primaryClass = "nucl-ex",
    doi = "10.1016/j.nuclphysa.2014.09.002",
    journal = "Nucl. Phys. A",
    volume = "931",
    pages = "103--112",
    year = "2014"
}

@article{Wang:2013duu,
    author = "Wang, Rui-qin and Shao, Feng-lan and Liang, Zuo-tang",
    title = "{Baryon-antibaryon flavor correlation in quark-combination models in heavy-ion collisions}",
    eprint = "1312.0185",
    archivePrefix = "arXiv",
    primaryClass = "hep-ph",
    doi = "10.1103/PhysRevC.90.017901",
    journal = "Phys. Rev. C",
    volume = "90",
    number = "1",
    pages = "017901",
    year = "2014"
}

@article{Poberezhnyuk:2019pxs,
    author = "Poberezhnyuk, R. and Vovchenko, V. and Motornenko, A. and Gorenstein, M. I. and Stoecker, H.",
    title = "{Chemical freeze-out conditions and fluctuations of conserved charges in heavy-ion collisions within quantum van der Waals model}",
    eprint = "1906.01954",
    archivePrefix = "arXiv",
    primaryClass = "hep-ph",
    doi = "10.1103/PhysRevC.100.054904",
    journal = "Phys. Rev. C",
    volume = "100",
    number = "5",
    pages = "054904",
    year = "2019"
}

@article{Gupta:2022phu,
    author = "Gupta, Sourendu and Mallick, Debasish and Mishra, Dipak Kumar and Mohanty, Bedangadas and Xu, Nu",
    title = "{Limits of thermalization in relativistic heavy ion collisions}",
    doi = "10.1016/j.physletb.2022.137021",
    journal = "Phys. Lett. B",
    volume = "829",
    pages = "137021",
    year = "2022"
}

@article{Karsch:2010ck,
    author = "Karsch, Frithjof and Redlich, Krzysztof",
    title = "{Probing freeze-out conditions in heavy ion collisions with moments of charge fluctuations}",
    eprint = "1007.2581",
    archivePrefix = "arXiv",
    primaryClass = "hep-ph",
    reportNumber = "CERN-PH-TH-2010-161",
    doi = "10.1016/j.physletb.2010.10.046",
    journal = "Phys. Lett. B",
    volume = "695",
    pages = "136--142",
    year = "2011"
}

@article{Fischer:2022oqp,
    author = "Fischer, H. G. and Makariev, M. and Varga, D. and Wenig, S.",
    title = "{A comprehensive study of the inclusive production of negative pions in p+p collisions for interaction energies from 3~GeV to 13~TeV covering the non-perturbative sector of the strong interaction}",
    eprint = "2202.09137",
    archivePrefix = "arXiv",
    primaryClass = "hep-ex",
    doi = "10.1140/epjc/s10052-022-10591-8",
    journal = "Eur. Phys. J. C",
    volume = "82",
    number = "10",
    pages = "875",
    year = "2022"
}

\end{document}